\documentclass[aps,prb,twocolumn,superscriptaddress,eqsecnum,nofootinbib,showpacs]{revtex4}
\pdfoutput=1
\usepackage{amssymb}
\usepackage{graphicx}
\usepackage{epsf,epsfig}
\usepackage{bm}
\usepackage{bbm}
\usepackage{amsfonts}
\usepackage{amsmath}
\usepackage[normalem]{ulem}

\newcommand{\be}{\begin{eqnarray}}
\newcommand{\ee}{\end{eqnarray}}
\newcommand{\ba}{\begin{array}}
\newcommand{\ea}{\end{array}}
\newcommand{\no}{\nonumber}
\newcommand{\tr}{\mbox{tr}}

\newcommand{\eps}{\varepsilon}

\newcommand{\bfr}{{\bf r}}
\newcommand{\bfrp}{{\bf r'}}

\newcommand{\bfn}{{\bf n}}

\newcommand{\bfp}{{\bf p}}

\newcommand{\bfj}{{\bf j}}

\newcommand{\bfq}{{\bf q}}

\newcommand{\bfg}{{\bf g}}

\newcommand{\bfrho}{{\bm \rho}}

\begin{document}

\title{Effective theory for the propagation of a wave-packet in a disordered and nonlinear medium}
\author{G. Schwiete}
\email{schwiete@zedat.fu-berlin.de} \affiliation{Dahlem Center for Complex Quantum Systems and Institut f\"ur Theoretische Physik,
Freie Universit\"at Berlin, Arnimallee 14, 14195 Berlin, Germany
}
\author{A. M. Finkel'stein}
\affiliation{Department of Physics and Astronomy, Texas A\&M University, College Station, TX 77843-4242, USA}
\affiliation{Department of Condensed Matter Physics, The Weizmann Institute of Science, 76100 Rehovot, Israel}
\date{\today}

\begin{abstract}
The propagation of a wave-packet in a nonlinear disordered medium exhibits interesting dynamics. Here, we present an analysis based on the nonlinear Schr\"odinger equation (Gross-Pitaevskii equation). This problem is directly connected to experiments on expanding Bose gases and to studies of transverse localization in nonlinear optical media. In a nonlinear medium the energy of the wave-packet is stored both in the kinetic and potential parts, and details of its propagation are to a large extent determined by the transfer from one form of energy to the other. A theory describing the evolution of the wave-packet has been formulated in [G.~Schwiete and A. Finkel'stein, Phys. Rev. Lett. {\bf 104}, 103904 (2010)] in terms of a nonlinear kinetic equation. In this paper, we present details of the derivation of the kinetic equation and of its analysis. As an important new ingredient we study interparticle-collisions induced by the nonlinearity and derive the corresponding collision integral. We restrict ourselves to the weakly nonlinear limit, for which disorder scattering is the dominant scattering mechanism. We find that in the special case of a white noise impurity potential the mean squared radius in a two-dimensional system scales linearly with $t$. This result has previously been obtained in the collisionless limit, but it also holds in the presence of collisions. Finally, we mention different mechanisms through which the nonlinearity may influence localization of the expanding wave-packet.
\end{abstract}

 \maketitle


\section{Introduction}
\label{sec:intro}

Currently, much attention is devoted to experiments studying the dynamics of a wave-packet evolving in the presence of both random scatterers and nonlinearity. These experiments are inspired by the idea that one can visualize the phenomenon of Anderson localization. The propagation of a wave-packet in the presence of multiple scattering on a random potential has been studied using photonic crystals\cite{Schwartz07,Lahini08} and also ultra-cold Bose gases confined initially inside a trap\cite{Clement05,Fort05,Lye05,Schulte05,Billy08,Roati08,Hulet09,Dries10,Robert10,Jendr12}. The nonlinearity in the case of photonics is induced by the Kerr effect (the change in the refractive index in response to an electric field), or may result from the particle-particle interactions in the case of cold atoms. In the optics experiments, a laser beam is sent into a nonlinear optical crystal with a refractive index varying randomly in the plane transversal to the direction of the pulse propagation. The resulting beam profile can be monitored on the opposite side of the crystal. In a second class of experiments, atoms condensed initially inside a trap are released and, during the subsequent expansion, are subjected to a disorder potential. Unlike in the case of photonic crystals, in the latter experiments it is possible to extract information about the full time-evolution of the expanding wave-packets.

Motivated by these experiments, we recently presented an effective theory of the propagation of a wave-packet (averaged over many disorder-realizations) injected in a disordered and nonlinear medium in two dimensions\cite{Schwiete10}. In the regimes preceding Anderson localization, or when it is absent, we found that the propagation of the wave-packet in a nonlinear disordered medium exhibits interesting dynamics related to the fact that in the presence of nonlinearities the energy of the wave-packet is stored both in the kinetic and potential parts. Then the propagation of the wave-packet is to a large extent determined by the transfer from one form of the energy to the other.

The derivation of the kinetic equation presented in Ref.~\onlinecite{Schwiete10} was based on a classical field theory, supplemented with the use of the quasiclassical approximation, a well-known tool in the theory of nonequilibrium superconductivity\cite{Eilenberger68,Larkin68,Kopnin01}. The corresponding functional can also be used as a basis for a diagrammatic perturbation theory. The relation between the different terms appearing in the kinetic equation and the diagrammatic perturbation theory was explained in Ref.~\onlinecite{Schwiete10a}. Recently, the kinetic equation was re-derived in Ref.~\onlinecite{Cherroret11} using a diagrammatic approach. In this article, we present details of the microscopic approach used for the derivation of the kinetic equation presented in Ref.~\onlinecite{Schwiete10}. We also include an important new ingredient into the formalism, inter-particle collisions. As a consequence, the resulting kinetic equation contains an additional term, the collision integral. We finally discuss the relevance of the collision processes.

We will assume that the time evolution of the injected wave-packet is governed by the nonlinear Schr\"odinger equation (NLSE), which is referred to as the Gross-Pitaevskii Equation (GPE) in the context of atomic Bose-Einstein condensates. The NLSE/GPE differs from the conventional Schr\"odinger equation by an additional cubic term (we set $\hbar=1$ for the GPE):
\be
&&i\partial_t\Psi({\bf r},t)\\
&=&-\frac{1}{2m}\nabla^2\Psi({\bf r},t)+u({\bf r})\Psi({\bf r},t)
+\lambda |\Psi({\bf r},t)|^2\Psi({\bf r},t).\no
\label{Eq:Gross}
\ee
For negative (positive) $\lambda$ the nonlinear term is of the self-focusing (de-focusing) type. This corresponds to an attractive (repulsive) potential $\lambda |\Psi({\bf r},t)|^2$.
The disorder potential $u({\bf r})$ is the source of randomness in the above equation. Starting from the NLSE/GPE, we derive a kinetic equation that describes the diffusive evolution of an injected wave-packet in a disordered nonlinear medium. Since the disorder we study is static, the kinetic equation preserves not only the integrated intensity/number of particles, but also the energy carried by the diffusing wave-packet. For a repulsive nonlinear term in the NLSE/GPE (that is typical for cold atoms), the potential energy stored in the medium is positive. Then, during the course of expansion, the potential part of the energy is gradually converted into the kinetic part, thereby increasing it. For an attractive nonlinearity (typical for optics), the potential energy stored in the medium is negative, and the dynamics is richer and may, in principle, include a collapse\cite{Vlasov71,Zakharov84,Sulem99}.

The NLSE used in optics is derived from the Maxwell equations using the so-called paraxial approximation, \cite{Shen84} and thus describes the evolution of the smooth envelope of the electric field. The propagation direction of the laser beam, say the $z$-direction, plays the role of time in the NLSE. In this sense, the disorder potential which results from random variations of the refractive index is static when it is $z$-independent (only such a system is considered here). For example, the two-dimensional ($2D$) transverse evolution of a pulse is studied in a $3D$ sample\cite{DeReadt89}.
In optics, the mass $m$ has to be replaced by the wave vector $k=\omega/c$, where $\omega$ is the frequency of the carrier wave and $c$ the velocity of light in the medium. The intensity of the beam is proportional to $|\Psi({\bf r},z)|^2$. We will be interested in the description of the wave-packet when its size $L=L(z)$ exceeds much the typical mean free path, $l_{\rm typ}$, which in turn is much larger than
the typical wave-length $\lambda_{typ} $ of the components constituting the wave
packet:
\be
L\gg l_{\rm typ}\gg\lambda_{\rm typ}.
\label{eq:L}
\ee
All three scales are related, of course, only to the propagation in the directions \emph{transverse} to $z$.

The GPE \cite{Gross61, Pitaevskii61} is commonly used for the description of a large ensemble of Bose-atoms confined inside a trap. We are, in turn, interested in the evolution of a cloud in which atoms are scattered by a random potential. The usage of the GPE in this context is worth commenting: The Schr\"odinger
equation for the field operators describing a many-body system, $\hat{\psi} (r,t)$, can be written as
\begin{equation}
i\partial_t\hat{\psi }=-\frac{1}{2m}\nabla^2 \hat{\psi }+u(\bfr)\hat{\psi }+\lambda\hat{\psi }^{\dagger }\hat{\psi }\hat{\psi },
\label{Eq:Gross operator}
\end{equation}
where, under the assumption that the scattering length $a_s$ is the shortest length in the problem, the potential of the particle interaction can be taken in the form $U(\bfr)=\lambda \delta (\bfr)$ (recall that for atoms $\lambda=4\pi\hbar^2a_s/m$, where $a_s$ is the scattering length). We will assume that occupation numbers $n_\bfp$ for the relevant momenta are large to ensure high occupancy. In this case the operators $\hat{\psi}$ in this equation may be substituted by a complex valued classical field $\Psi$ (for a formal discussion of this point see, e.g., Ref.~\onlinecite{Kagan97}).
It is worth mentioning that in the case of quantum electrodynamics a similar step leads to the classical Maxwell equations for large photon occupation numbers. It will be important for us that the field $\Psi(\bfr,t)$ should not necessarily be interpreted as a condensate wave function in order to be described by the GPE. The density of the cloud can be expressed as $|\Psi (\bfr,t)|^2$.

In addition to the condition of Eq.~(\ref{eq:L}), throughout this paper it will be assumed that
\be
\lambda_{\rm typ}\gg a\gg\ a_s,
\ee
where $a$ is the inter-particle distance of atoms in the cloud. The former inequality corresponds to a high occupancy of atoms which justifies the use of the classical GPE for the description of the Bose gas. The latter inequality means (by definition) that the gas is dilute. Since we study the effects of the nonlinearity, we are nevertheless interested in a situation for which the gas is sufficiently dense in the sense that the energy per atom induced by the nonlinearity, which is of the order of $\lambda |\Psi(
\bfr,t)|^2$, is not negligible compared to the typical kinetic energy of the atoms constituting the cloud.

In line with most of recent experiments on cold atoms/photonic crystals, we will study the density/intensity averaged over many realizations of disorder. Correspondingly, we are interested in the evolution of the wave-packet on length scales exceeding the typical mean free path $l_{typ}$. To obtain an averaged description for the propagation of the cloud, one needs to introduce the smooth disorder averaged density, ${n}(\bfr,t)=\left\langle |\Psi(\bfr,t)|\right\rangle_{dis}$. As a result, the nonlinearity generates a term of the form $2\lambda n(\bfr,t) \Psi (\bfr,t)$, i.e., it gives rise to a self-consistent potential $\vartheta(\bfr,t)=2\lambda n(\bfr,t)$. We would like to stress that while the density $n(\bfr,t)$ is smooth on the scale of the mean free path, the wave function $\Psi(\bfr,t)$ is not. Indeed, in the case we study the wave function varies rapidly on this scale, since the wavelength is assumed to be much smaller than the mean free path.  A similar-looking term, $2\lambda n(\bfr,t) \Psi(\bfr,t)$, arises in the description of a coupled system of condensate and non-condensate particles, where $n$ stands for the density of non-condensate particles, while $\Psi$ is the smooth condensate wave function.\cite{Griffin09, Kamenev11} In contrast, in our description $n$ is the density of the \textit{whole} gas.

The self-consistent potential is not the only effect originating from the nonlinearity that contributes to the effective kinetic theory of wave-packet propagation. Indeed, in the next order in the nonlinearity $\lambda$, the so-called collision integral arises, which describes inter-particle collisions. We will discuss this issue for atoms for which the meaning of collisions is more obvious. To get an idea about the collision rate, let us first consider the rate of two-body collisions in the gas of small density, for which the occupation numbers are small, $n_\bfp \ll 1$. In the three-dimensional case, the collision rate is the inverse of the Maxwell-Boltzmann collision time: $1/\tau_{MB}=\sqrt{2}n(r)\sigma v_{\varepsilon}$, where the atomic cross section $\sigma=8\pi a_{s}^{2}$ and $v_{\varepsilon}$ is the velocity of a particle with the energy $\varepsilon$. Then, $1/(\tau _{MB}\varepsilon)\sim (a_s/a)^2(\lambda_{\eps}/a)$, which in a dilute gas with small occupation numbers is a product of two small factors ($\lambda_{\eps}$ is the wave-length of a particle with the energy $\eps$). The situation changes radically for a gas with large occupation numbers, $n_{\bfp}\gg 1$. The smallness induced by the scattering length $a_s$ in the dilute gas, can be compensated by large factors $n_{\bfp}$. (The balancing between the smallness of the interaction amplitude and large occupation numbers is specific for Bose-gases as
compared to fermionic systems.) As a result, one gets for the collision rate $1/\tau_{coll}\sim \lambda_{\overline{\eps}}^2 {n^2}/{\overline\eps}$, where $\overline{\eps}$ is a typical kinetic energy of the Bose-atoms. Let us finally emphasize that while we used here the language appropriate for atomic gases, the collision rate $1/\tau_{coll}$ has its origin in the nonlinearity and as such this estimate is relevant for any system described by the NLSE/GPE irrespective of its microscopic origin.

The kinetic equation presented in this paper is derived for the case when disorder is responsible for the dominant scattering mechanism, $1/\tau \gg 1/\tau_{coll}$. To be in correspondence with this inequality, we will limit
ourself to the case when the effect of nonlinearity is sufficiently weak so that $\lambda n(\bfr)\ll \overline{\eps}(\bfr)$.

It is worth commenting on an important byproduct of the interaction smallness. Under the condition $\lambda n(\bfr)\ll \overline{\eps}(\bfr)$ we need not consider the transition to the Bogoliubov spectrum. This is because under this condition only a tiny fraction of the states with the smallest energies is influenced by the off-diagonal components in the Bogoliubov Hamiltonian. For the majority of the particles the off-diagonal components of the Bogoliubov Hamiltonian can safely be ignored.

As it was already mentioned, when treating disorder we assume that the mean free path is much larger than the typical wavelength $\lambda_{typ}$ of the components constituting the wave-packet. Throughout this paper, we use the model of a delta-correlated Gaussian disorder potential, characterized by $\left\langle u(\bfr)u(\bfr')\right\rangle=\gamma\delta(\bfr-\bfr')$. This model is appropriate if scattering occurs on quantum impurities, for which the range of the potential is much smaller than the wavelength $\lambda_{typ}$.  For the delta-correlated disorder potential, the density of states determines the frequency-dependence of the scattering rate, $1/\tau(\eps)=2\pi \nu(\eps)\gamma \propto \eps^{(d-2)/2}$ and of the diffusion coefficient $D(\eps)=2\eps\tau(\eps)/md\propto \eps^{2-d/2}$. In particular, the scattering rate in $d=2$ is energy-independent. Both in optics experiments and in experiments on Bose gases, one often uses speckles to realize the disorder potential. The speckle potential has a finite correlation length. If the wave-length $\lambda_{typ}$ is much larger than the correlation length, the model of the delta-correlated disorder potential remains a good approximation. If the wavelength is sufficiently short to resolve the finite correlation length, however, one needs to be more cautious.
Unlike for the short range scatterers, the typical time for the randomization of the momentum direction, i.e., the transport scattering time, no longer coincides with the single particle scattering time, which is determined by the imaginary part of the self-energy in the disorder averaged Green's function. The transport scattering time $\tau_{tr}(\eps)$ acquires a frequency dependence that differs from the one for short range scatters stated above. The same is true for the diffusion coefficient, since it depends on $\tau_{tr}$ as $D= 2\eps \tau_{tr}(\eps)/md$. The expression for $\tau_{tr}$ appropriate for a speckle potential can be found in the literature, e.g., in Refs.~\onlinecite{Kuhn07}. As concerns the nonlinear diffusion equation derived in this manuscript, it can be expected that the only change that needs to be introduced when dealing with a speckle potential is the replacement $\tau\rightarrow \tau_{tr}$ in the final form of the equations, which already contains a energy-dependent diffusion coefficient.

The paper is organized as follows. In Sec.~\ref{Sec:discussion} we proceed directly to the discussion of the
nonlinear kinetic equation. Those readers, who are not interested in the technical details of the derivation of the kinetic equation based on the quasiclassical approximation, find the most important information in Sec.~\ref{Sec:discussion} as well as in the Conclusion. First, we discuss the equation in the collisionless regime in Sec.~\ref{Sec:discussion_collisionless}. Although most of the material of Sec.~\ref{Sec:discussion_collisionless} has already been presented in Ref.~\onlinecite{Schwiete10}, we include it here in order to make the paper self-contained. In the second part, Sec.~\ref{Sec:discussion_collisions}, we add the effect of collisions. It turns out that the interparticle collisions impose certain constraints on the range of validity of the derived equations. The main result of this paper is formulated here:  In two spatial dimensions, the mean squared radius of the wave-packet grows linearly in time. This result is not affected by inter-particle collisions.

In Sec.~\ref{Sec:basicformalism} we introduce the field theory approach that is the main tool for our investigations. The basic idea is to write a functional integral expression for the time evolution of the observable in question. (Our aim here is to describe the evolution of the density/intensity $n=|\Psi|^2$. The wave function at the initial time $\Psi_0$ is assumed to be known.) Typically, this kind of approach is used when studying Langevin-type equations including a noise term with a given correlation function. In the problem under study in this paper, no noise is considered. Instead, we use an analogous construction, and then average over disorder configurations.
The resulting theory closely resembles the structure one encounters in Keldysh field theories, where Green's functions can be transformed to a block-triangular form. Retarded and advanced Green's functions are supplemented by a third type of Green's function that contains information about the distribution function $n(\bfr,t,\eps)$, which we are interested in.

In Sec.~\ref{Sec:timeevolution} the averaging over the disorder potential is performed, i.e., we provide a description of the evolution of a wave-packet averaged over many disorder configurations (realizations). First, the theory of the wave-packet in the absence of the nonlinearity is discussed. Here, we make contact with Ref.~\onlinecite{Shapiro07} and \onlinecite{Shapiro12}, where the expansion of a Bose-condensate over a disorder potential was studied starting from a later stage of the time-evolution when the nonlinearity may already be neglected.  Afterwards, the nonlinear problem is considered. We start this discussion with a diagrammatic analysis (in two dimensions) before deriving the kinetic equation using the method of quasiclassical Green's functions. Here, we proceed in close analogy with the theory of nonhomogeneous superconductivity \cite{Eilenberger68,Larkin68,Kopnin01}. The main result of Sec.~\ref{Sec:timeevolution} is given by Eq.~(\ref{Eq:basic22}), which is a classical nonlinear diffusion equation in the collisionless regime. The equation was first presented and analyzed in Ref.~\onlinecite{Schwiete10} for a two-dimensional system. Discussion of two dimensions was of special interest for us, because for weak disorder there is an exponentially large diffusive regime before the Anderson localization takes place. After our work\cite{Schwiete10}, the equation Eq.~(\ref{Eq:basic22}) was re-derived and generalized for arbitrary dimensions in Ref.~\onlinecite{Cherroret11}, using the diagrammatic technique. It was noted that for a generalization to dimensions $d\ne 2$ a new term in the kinetic equation is required in order to account for the non-constancy of the density of states. In Sec.~\ref{Sec:timeevolution} the equation is obtained for arbitrary dimensions $d=2,3$ including the additional term found in Ref.~\onlinecite{Cherroret11}.

In Sec.~\ref{Sec:collisions} we derive the collision integral in the kinetic equation originating from the NLSE/GPE. We provide a diagrammatic interpretation of the different terms contributing to the collision integral. Finally, we conclude in Sec.~\ref{Sec:conclusion} with a discussion of the results. In particular, we comment on the role of the nonlinearity in the context of localization.

\section{Discussion of the nonlinear kinetic equation}

\label{Sec:discussion}

\subsection{The kinetic equation in the collisionless regime}
\label{Sec:discussion_collisionless}

We start from the nonlinear kinetic equation determining the density evolution in the diffusive regime. The argument $\tilde{\varepsilon}$ in this equation has the physical meaning of the kinetic energy,
$\tilde{\varepsilon}({\bf r},t)=\epsilon-\vartheta({\bf r},t)$, while  $\vartheta({\bf r},t)$ is a self-consistent potential. Correspondingly, the diffusion coefficient is ${D}_{\tilde{\eps}}=\tilde{v}_{\tilde{\eps}}^2\tau_{\tilde{\eps}}/d$. Then, the equation for the distribution function looks as follows:
\be
&&\partial_t\tilde{n}(\bfr,t,\eps)-\nabla(D_{\tilde{\eps}}{\nabla}_{\Gamma}\tilde{n}(\bfr,t,\eps))\no\\
&&+\partial_t\vartheta(\bfr,t)\partial_\eps\tilde{n}(\bfr,t,\eps)=\delta(t)2\pi\nu(\tilde{\eps}) F(\tilde{\eps},\bfr).\label{eq:fundam}
\ee
This equation should be supplemented with the self-consistency relation for the potential $\vartheta({\bf r},t)=2\lambda n({\bf r},t)$, where
\be
n({\bf r},t)=\int \frac{d\varepsilon}{2\pi}\; \tilde{n}({\bf r},t,\varepsilon).
\ee
Note that the diffusion term contains a sort of the covariant derivative:
\be
\nabla_\Gamma=\nabla-\nabla\vartheta(\bfr,t)\Gamma_{\tilde{\eps}},
\ee
where $\Gamma_{\tilde{\eps}}=-\partial_\eps\ln \nu(\tilde{\eps})$.
The term on the right-hand side of Eq. (\ref{eq:fundam}) specifies the injection of the wave-packet and initial evolution up to times of the order of the scattering time $\tau$. Namely,
\be
F(\varepsilon,{\bf r})=\int \frac{d\bfp d\bfq}{(2\pi)^{2d}}\;F({\bf p},{\bf q})\;\mbox{e}^{i{\bf q}{\bf r}}\;2\pi \delta(\varepsilon-\varepsilon_{\bf p}),
\ee
and $F({\bf p},{\bf q})=\Psi_0({\bf p}+{\bf q}/2)\Psi_0^*({\bf p}-{\bf q}/2)$ is determined by the initial wave function $\Psi_0$. Further, $\varepsilon_{\bf p}=p^2/(2m)$ is the kinetic energy.

Despite its apparent simplicity, Eq. (\ref{eq:fundam}) is a rather complicated nonlinear integro-differential equation. The diagrammatic interpretation of the different terms appearing in this equation is provided in Sec.~\ref{sec:intperturbation} for the two-dimensional case. The main new ingredient for $d\ne 2$ is the non-constant density of states, $\nu(\tilde{\eps})$. (Note that $\Gamma$ vanishes in two spatial dimensions when the density of states is constant. In three dimensions, however, $\Gamma=\frac{2-d}{2\eps}$ is finite.) Since the density of states enters with the argument $\tilde{\eps}=\eps-\vartheta(\bfr,t)$, the scattering rate acquires an explicit dependence on $\vartheta$. This eventually leads to a modification of the diffusion term in Eq. (\ref{eq:fundam}) by substituting $\nabla\rightarrow\nabla_\Gamma$, which was first noticed in Ref.~\onlinecite{Cherroret11}.

The underlying physics of the nonlinear diffusion equation Eq.~(\ref{eq:fundam}) was discussed in Ref.~\onlinecite{Schwiete10}. The equation describes diffusion of a particle with total energy $\eps$ on the background of a smoothly varying potential $\vartheta$. Correspondingly, the kinetic energy $\varepsilon_{\bfp}=\eps-\vartheta$ varies locally in space and time. One may notice that in the NLSE/GPE a purely time dependent potential may be removed by a gauge transformation $\Psi({\bf r},t)\rightarrow \Psi({\bf r},t)\exp\left(-i\int_{t_0}^t dt' V(t')\right)$, that leaves the density $|\Psi({\bf r},t)|^2$ unchanged. On the level of the discussed equation, this point becomes obvious when writing the distribution function as a function of the kinetic energy instead of the total one, $n({\bf r},\varepsilon,t)=\tilde{n}({\bf r},\varepsilon+\vartheta({\bf r},t),t)$. Expressed in the new coordinates the equation reads
\be
&&\partial_{t}n({\bf r},\varepsilon,t)\no\\
&&-\Big[\nabla-\nabla \vartheta({\bf r},t)\partial_{\varepsilon}\Big]D_{\varepsilon}\Big[\nabla_\Gamma-\nabla\vartheta({\bf r},t)\partial_{\varepsilon}\Big]n({\bf r},\varepsilon,t)\no\\
&&=\delta(t)\;F(\varepsilon,{\bf r}),\label{Eq:basic22}
\ee
where now $\nabla_\Gamma=\nabla-\nabla\vartheta(\bfr,t)\Gamma_{\eps}$.
One can see explicitly that a purely time-dependent potential drops from the equation since $\vartheta({\bf r},t)$ enters only in combination with a spatial derivative, as $\nabla \vartheta(\bfr,t)$.
In Eq.~(\ref{Eq:basic22}), the diffusion coefficient $D(\eps)=2\eps \tau(\eps)/md$ depends explicitly on $\eps$, but also implicitly through $\tau(\eps)$. Within our model of a delta-correlated impurity potential, the elastic scattering rate $1/\tau(\eps)$ acquires a frequency dependence through $\nu(\eps)$. The form of the equation suggests, however, that it will also hold in the case of impurity potentials with a finite correlation length, when $\tau(\eps)$ should be replaced by the transport scattering time $\tau_{tr}(\eps)$.

It seems clear that a closed form solution of the nonlinear equation for arbitrary initial conditions cannot be found. In order to make progress we will rely on the use of conservation laws. The GPE describes a system in which the total particle number (or intensity in the case of the NLSE) and the total energy are conserved. The total momentum is not conserved, since the disorder potential breaks translational invariance. It is important to check that our approximations are consistent with the conservation laws, namely that energy and number conservation are still encoded in the nonlinear diffusion equation (\ref{Eq:basic22}).

Let us start with the number conservation. For that, we integrate Eq.~(\ref{Eq:basic22}) in $\eps$ and obtain the continuity equation in the form  $\partial_tn(\bfr,t)+\nabla{\bf j}(\bfr,t)=\delta(t)n(\bfr, t)$. The role of the right hand side is merely to determine the boundary condition at the initial time $t=0$. The expression for the current is
\be
{\bf j}(\bfr,t)&=&\int\frac{d\eps}{2\pi} \bfj(\bfr,\eps,t)\\
{\bf j}(\bfr,\eps,t)&=&-D_\eps[\nabla_\Gamma-\nabla\vartheta\partial_\eps]n(\bfr,\eps,t)
\ee

Next we turn to energy conservation. Here, the continuity equation, $\partial_t\rho_E(\bfr,t)=-\nabla{\bf j}_E$, takes the following form:
\be
\rho_E(\bfr,t)&=&\overline{\eps}(\bfr,t)+\lambda n^2(\bfr,t)\\
{\bf j}_E(\bfr,t)&=&\int\frac{d\eps}{2\pi} (\eps+\vartheta){\bfj}(\bfr,t,\eps)
\ee
where $\overline{\eps}=\int (d\eps/2\pi)\; \eps n(\bfr,t,\eps)$ can be interpreted as the average kinetic energy. In particular, we may conclude that the total energy
\be
E_{tot}=\int d{\bf r}\; (\overline{\varepsilon}+\lambda n^2)
\ee
is conserved. The total energy is conserved for our problem, because impurity scattering is elastic and we consider a closed system. The conservation of energy is a known property of NLSE/GPE from which we started. The derivation based on the kinetic equation, which we presented here, can be regarded as a check of the validity of our approach.

Remarkably, as we have observed in Ref.~\onlinecite{Schwiete10}, for two spatial dimensions when $\Gamma=0$, and if the scattering time is frequency-independent, the conservation laws completely determine the time evolution of the mean radius squared of the wave-packet, $\left\langle r_t^2 \right\rangle \equiv \int d{\bf r}\;r^2\;n({\bf r},t)/N$. Indeed,
in $2d$ the expression for the current ${\bf j}(\bfr,t)$ can be simplified and the continuity equation takes the form
\be
\partial_tn(\bfr,t){-\frac{\tau}{m}\nabla^2(\overline{\eps}+\lambda n^2(\bfr,t))}=\delta(t)n(\bfr, t)\label{Eq:basic33}.
\ee
Now multiplying Eq.~(\ref{Eq:basic33}) by ${r}^2$ and subsequently integrating in ${\bf r}$ one obtains that
\be
\partial_t\left\langle{r}_t^2\right\rangle =4D_{\varepsilon_{tot}},\label{eq:dtrsq}
\ee
where $\varepsilon_{tot}=E_{tot}/N$. The linear dependence of the mean square radius on time during the evolution is guarded by energy conservation. This is a rather non-trivial result; the rate of expansion is proportional not to $D_{\overline{\eps}}$, as one may naively expect, but to $D_{E_{tot}}$. The reason is that the rate of expansion combines the effect of diffusion and propagation in the field of the force induced by the self-consistent potential. This is one of the central results of our previous paper\cite{Schwiete10}; unfortunately, in higher dimensions it seizes to be valid due to the non-constancy of the density of states.

It remains to discuss general features of wave-packet dynamics in the repulsive and the attractive case.
When the potential energy related to the nonlinearity is converted into kinetic energy, the total kinetic energy increases in the repulsive case and decreases in the attractive case. Correspondingly, during the course of the expansion localization effects can be expected to be weakened for repulsive nonlinearity and enhanced for attractive nonlinearity. In particular, for an attractive nonlinearity the slowing down and eventual localization of the injected pulse (not considered here) occurs at smaller distances than in the linear case as observed in the experiment \cite{Schwartz07}. As it was indicated in Ref.~\onlinecite{Schwiete10}, if a part of the cloud lags behind, this fragment may have a strong tendency to localize. One may expect that this kind of localized fragment generically remains from an expanding cloud when the nonlinearity is attractive.
To check this point, it would be desirable to analyze data with respect to the intensity/number of particles of the remaining localized part of the cloud and, if possible, the energy concentrated in this part as compared to that in the initial cloud.

\subsection{The role of collisions}
\label{Sec:discussion_collisions}

The nonlinear term in the NLSE/GPE gives rise to a collision integral in the diffusive kinetic equation, which is proportional to $\lambda^2$. The full kinetic equation including interparticle collisions takes the form
\be
&&\partial_{t}n({\bf r},t,\eps)\no\\
&&-\Big[\nabla-\nabla \vartheta({\bf r},t)\partial_{\varepsilon}\Big]D_{\varepsilon}\Big[\nabla_\Gamma-\nabla\vartheta({\bf r},t)\partial_{\varepsilon}\Big]n({\bf r},t,\eps)\no\\
&&=\delta(t)\;F({\bf r},\eps )+2\pi\nu(\eps) I^{\rm coll}(\bfr,t,\eps)
\label{Eq:Fullequation}
\ee
with
\be
&&I^{\rm coll}(\bfr,t,\eps)\no\\
&&=4\pi \lambda^2 (2\pi)^d \int d\bfn  d\bfn_2d\bfn_3d\bfn_4\int  d\eps_2 d\eps_3 d\eps_4 \no\\
&&\times \nu(\eps_2) \nu(\eps_3)\nu(\eps_4) \delta(\eps+\eps_2-\eps_3-\eps_4)\no\\
&&\times \delta(\bfp_\eps +\bfp_{\eps_2}-\bfp_{\eps_3}-\bfp_{\eps_4})\Big([n'_{\eps}+n'_{\eps_2}]n'_{\eps_3}n'_{\eps_4}\no\\
&&-n'_{\eps}n'_{\eps_2}[n'_{\eps_3}+n'_{\eps_4}]\Big),
\label{Eq:CollisionInt}
\ee
where $2\pi\nu(\eps) n'_\eps(\bfr,t)=n(\bfr,t,\eps)$, $\bfp_\eps=p_\eps\bfn$, $\bfp_{\eps_i}=p_{\eps_i}\bfn_i$, and the integration goes over positive frequencies only.
To conclude, we get a standard collision term of two particles in the limit of large occupation numbers $n_{\eps_i}'\gg 1$. The left-hand side of the kinetic equation takes into consideration that the distribution function of states participating in the collision are determined by the diffusive propagation in the disordered and nonlinear medium.

The collision integral contains two terms describing the "in"- and "out"-collision channels. To estimate the scattering rate $1/\tau_{coll}$, let us focus on the "out"- term, which is given by the last term in the expression for $I_{coll}$, Eq.~(\ref{Eq:CollisionInt}), and is proportional to $n'_{\eps}$. We will write it as $n_{\eps}/\tau_{coll}$. Recall that the typical kinetic energy per particle at point $\bfr$ is denoted as $\overline{\eps}(\bfr)$. For a conservative estimate of the scattering rate, let us consider an energy $\eps\sim\overline{\eps}(\bfr)$; in this case the kinematic constraints induced by the momentum and energy conservation in the collision integral are minimal. Since one has to integrate two distribution functions over energies, this ultimately yields a factor $n^2(\bfr)$. As a result one gets
\be
\frac{1}{\tau_{coll}}\sim \lambda^2 \frac{n^2(\bfr)}{\overline{\eps}(\bfr)}.
\ee
It is clear from this estimate that in order to use the language of the kinetic equation with well defined distribution function $n({\bf r},t,\eps)$, one has to be limited to the case when $\lambda n(r,t)\ll\overline{\eps}(\bfr)$. Under this condition, $1/\tau_{coll}\ll\overline{\eps}(\bfr)$.

Still, there remains a question about a comparison of the rate of inter-particle collisions with elastic scattering caused by disorder, i.e., $1/\tau_{coll}$ versus $1/\tau$. In this paper we limit ourself to the case of rare collisions, $1/\tau_{coll}\ll 1/\tau$, i.e., we assume that elastic scattering events occur more frequently than inter-particle collisions. This condition is more restrictive than the condition $\overline{\eps}(\bfr)\gg 1/\tau_{coll}$ discussed above.

The collisions, naturally, change the dynamics of the propagation. As long as the kinetic equation in the derived form holds, however, the result (\ref{eq:dtrsq}) about the rate of the expansion of the wave-packet remains valid even in spite of the inter-particle collisions. This is because (i) the collision integral is local and as such does not change the mean radius squared of the wave-packet, $\left \langle r^2\right\rangle_{dis}$.
Furthermore, (ii) in two spatial dimensions, the rate of expansion depends only on the total energy $E_{tot}$, which is not altered by collisions and it does not depend on the energy dependence of the distribution function, which is controlled by the collision integral.

Finally, we would like to note that while the rate of "delivery" of colliding particles was controlled by diffusion, we did not consider the modifications of the collision integral by disorder. It is is very different from what happens in disordered conductors at low temperatures, $T\ll 1/\tau\ll\eps_F$. The reason is that the kinetics of the classical particles, not constrained by the existence of the Fermi-surface, is similar to the case for which $1/\tau\ll T\sim \eps_F$ with $\eps_F\sim \overline{\eps}$, where $\eps_F$ is the Fermi energy. Then, modification of the collision integral by disorder leads to a smallness $1/\tau\overline{\eps}(\bfr)$ without gaining a large factor $1/(\tau T)$, as it was in the case of conductors at low temperature.

\section{Basic formalism}
\label{Sec:basicformalism}
In this section we introduce the field theory approach that is the tool for our investigations. Our aim is to describe the evolution of the density (intensity) $n=|\Psi|^2$, \emph{averaged over disorder configurations}. The wave function at the initial time $\Psi_0$ is assumed to be known.

Formally, the problem bears a certain similarity with the description of critical dynamics near a phase transition, or, more generally, the study of Langevin-type equations with the help of field theory approaches. The formalism we are alluding to here is often called Martin-Siggia-Rose (MSR) formalism \cite{Martin73,Janssen76, DeDominicis78, Kamenev11} and finds applications in many different branches of physics. The basic idea is to write a functional integral expression for the time evolution of the observable in question. With the help of a delta-function entering the integral, the wave function is fixed to coincide with the solution of the underlying equation. By introducing an additional field variable and thereby doubling the degrees of freedom, one may write the delta-function with the help of an integral over an exponentiated action.

Typically, this kind of approach is used when studying dynamical problems, for which the original equation contains a noise term with known correlation function. One may then average over the noise, and study the resulting functional with field theoretical methods like perturbation theory, the renormalization group, or by analyzing instantonic configurations. In the problem under study in this paper, no noise is present. Instead, we use an analogous construction, and then average over disorder configurations. With a proper regularization, vacuum loops are absent right from the beginning and this is why the dynamical approach is particularly useful for the problem of quenched disorder, as was already noted long time back \cite{DeDominicis78}.

The resulting field theory indeed closely resembles the structure one encounters in Keldysh field theories, where Green's functions can be transformed to a block-triangular form. Retarded and advanced Green's functions are supplemented by a third type of Green's function that contains information about level population.

For a Bose-Einstein condensate, one can obtain the Gross-Pitaevskii equation as a mean field equation for the full quantum many-body problem. As one might expect from this observation, a connection exists between Keldysh-type field theories for the quantum problem, and the MSR-type approach. Indeed, in the Keldysh approach, two distinct types of interaction vertices exist, they are sometimes referred to as  quantum and classical vertices\cite{Kamenev11}.  By disregarding the quantum vertices, while retaining the classical ones, one recovers a representation of the functional delta-function, that fixes the evolution of the (classical) fields to obey the classical equation of motion, in this case the Gross-Pitaevskii equation. This approach additionally allows to consider correlations in the initial density matrix, and one can obtain, for example, the so-called truncated Wigner approximation, as explained in more detail in Ref.~\onlinecite{Polkovnikov03}. In optics, the nonlinear Schr\"odinger equation emerges as a result of the paraxial approximation applied to the Helmholtz-equation \cite{Shen84} and has thus a different microscopic origin. This is the reason why we do not explicitly use the (microscopic) Keldysh approach as a starting point in this paper.

\subsection{Action}
Our starting point is the Gross-Pitaevskii equation in the form given in Eq.~({\ref{Eq:Gross}}). This equation describes the time evolution of the macroscopic wave-function $\Psi(\bfr,t)$ in the presence of an external potential $u(\bfr)$. The total density $|\Psi(\bfr,t)|^2$ is conserved in time and we use the normalization $\int d{\bf r} |\Psi({\bf r},t)|^2=N$, where $N$ is the total number of atoms in the gas. The quantity of our interest is the disorder averaged density
\be
n({\bf r},t)=\left\langle |\Psi({\bf r},t)|^2\right\rangle_{dis}.
\ee
Disorder averaging $\left\langle\dots\right\rangle_{dis}$ is performed with the help of the Gaussian probability distribution
\be
\mathcal{P}(u)=\mathcal{N}\exp\left(-\frac{1}{2} \int d\bfr \;u(\bfr)W^{-1}(\bfr-\bfr')u(\bfr')\right),\label{eq:distribution}
\ee
where $\mathcal{N}$ provides the normalization, so that $\int Du \;\mathcal{P}(u)=1$. This definition implies that $\left\langle u(\bfr)\right\rangle_{dis}=0$ and $\left\langle u(\bfr)u(\bfr')\right\rangle=W(\bfr-\bfr')$. In this paper, we consider the specific case of a delta-correlated (white noise) potential, for which $W(\bfr-\bfr')=\gamma\delta(\bfr-\bfr')$. In two dimensions the density of states $\nu$ is constant, and one can identify $\gamma={1}/{2\pi\nu \tau}$, where $\tau$ is the scattering time.

We first note that the unaveraged density can be represented as the following functional average:
\be
n(\bfr,t)=\int D(\psi,\psi^*) D(\eta,\eta^*)\;|\Psi(\bfr,t)|^2\;\mbox{e}^{iS},
\ee
where we introduced the complex fields $\eta$, $\eta^*$ and $\psi$, $\psi^*$. The action $S$ is given by
\be
S&=&\int d\bfr d\bfr' dt dt'\;\left[\ba{cc}\psi^*(\bfr,t)\\\eta^*(\bfr,t)\ea\right]^T\tilde{g}^{-1}(\bfr,\bfr',t,t')\left[\ba{cc}\psi(\bfr,t)\\\eta(\bfr,t)\ea\right]\no\\
&&+i\int d\bfr\;[\eta(\bfr,0)\Psi_0^*(\bfr)-\eta^*(\bfr,0)\Psi_0(\bfr)],
\ee
where the inverse matrix Green's function $\tilde{g}^{-1}$ has the structure
\be
\tilde{g}^{-1}=\left(\ba{cc}0 &\tilde{g}^{-1}_A\\\tilde{g}_R^{-1}&0\ea\right).
\ee
The retarded and advanced Green's functions $\tilde{g}_{R/A}$ fulfill the equation
\be
&&\left(i\partial_t+\frac{\nabla^2}{2m}-u(\bfr)-\lambda|\psi(\bfr,t)|^2\right)\tilde{g}_{R/A}(\bfr,\bfr',t,t')\no\\
&&=\delta(\bfr-\bfr')\delta(t-t')
\ee
with standard boundary conditions. Indeed, upon integration in the auxiliary fields $\eta(\bfr,t)$, $\eta^*(\bfr,t)$ one obtains a functional delta function that fixes the fields $\Psi(\bfr,t)$ and $\Psi^*(\bfr,t)$ to obey the Gross-Pitaevskii equation and its complex conjugate, respectively. The last part of the action involving the fields $\Psi_0$ and $\Psi_0^*$ fixes the boundary conditions at the initial time, $\Psi(\bfr,t_0)=\Psi_0(\bfr)$ and $\Psi^*(\bfr,t_0)=\Psi^*_0(\bfr)$.  We see that the formalism involves a doubling of the degrees of freedom, similar to the Keldysh or closed-time-path approaches for quantum systems \cite{Kamenev11}, where two fields are introduced on forward and backward time-contours. We repeat that with a proper regularization vacuum loops are absent. For a more detailed account of the construction of the classical functional and the appropriate regularization we refer to Refs.~\cite{Jeon05,Berges07,Kamenev11}

In order to lighten the notation, we find it convenient to introduce the field doublets $\phi=\left(\psi,\eta\right)^T$ and $\overline{\phi}=\phi^\dagger\sigma_x=(\eta^*,\psi^*)$, so that
\be
n({\bfr},t)&=&\left\langle\tr\left(\sigma_-\left[\phi(\bfr,t)\otimes\overline{\phi}(\bfr,t)\right]\right)\right\rangle \label{eq:nfromfunctional}
\ee
Pauli matrices $\sigma_i$ act in the space of the the fields $\psi$ and $\eta$, and $\sigma_-=(\sigma_x-i\sigma_y)/2$. The averaging $\left\langle \dots\right\rangle=\int D(\phi,\phi^\dagger) \;(\dots) \exp(iS)$ is performed with respect to the action $S$, which we write in terms of $\phi$ and $\overline{\phi}$ and split into several parts,
\be
S=S_0+S_{s}+S'_{dis}+S_{int}.
\ee
The term $S_0$ alone describes the free propagation of fields $\phi$ in the absence of interactions and impurities. The source term $S_s$ contains information about the initial conditions, for convenience we choose $t_0=0$ from now on.
The disorder potential and the nonlinear term in the Gross-Pitaevskii equation give rise to $S'_{dis}$ and $S_{int}$, respectively.
\be
S_{0}&=&\int d\bfr d\bfrp dt
dt'\;\overline{\phi}(\bfr,t){g}_0^{-1}(\bfr-\bfr',t-t')\phi(\bfrp,t'),\\\no\\
S_{s}&=&i\int d\bfr\;\left(\overline{\phi}_0(\bfr)\phi(\bfr,0)-\overline{\phi}(\bfr,0)\phi_0(\bfr)\right),\\
S_{dis}'&=&-\int d\bfr dt \;\overline{\phi}(\bfr,t)u(\bfr)\phi(\bfr,t),\\
S_{int}&=&-\lambda\int d\bfr dt \;\overline{\phi}(\bfr,t)\phi(\bfr,t)\;\overline{\phi}(\bfr,t)\sigma_{-}\phi(\bfr,t).\label{eq:sint}
\ee
Here, the $2\times 2$ matrix Green's function
\be
g_0=\left(\ba{cc}g_0^R&0\\0&g_0^A\ea\right)\label{eq:g_0}
\ee
is composed of the retarded and advanced Green's functions $g_0^{R/A}(\bfp,\eps)=(\eps-\eps_\bfp\pm i\delta)^{-1}$, where $\eps_\bfp=\bfp^2/2m$.
For the initial condition, we introduced
\be
\phi_0(\bfr)=\left(\Psi_0(\bfr),0\right)^T,\quad  \overline{\phi}_0(\bfr)=\left(0,\Psi^*_0({\bf r})\right).
\ee

It is an important consequence of the structure of the theory, that the Green's function $G=-i\left\langle \phi\overline{\phi}\right\rangle$ has a triangular structure, where in accord with Eq.~(\ref{eq:g_0}) the $11$ and $22$ elements are retarded and advanced Green's functions. These Green's functions contain information about the spectrum, while the off-diagonal (12) element contains information about the occupation, in analogy to the Keldysh approach. Importantly, the $21$-element is equal to zero.

\subsection{Diagrammatic representation}

\begin{figure}[h]
\setlength{\unitlength}{2.3em}
\includegraphics[width=10\unitlength]{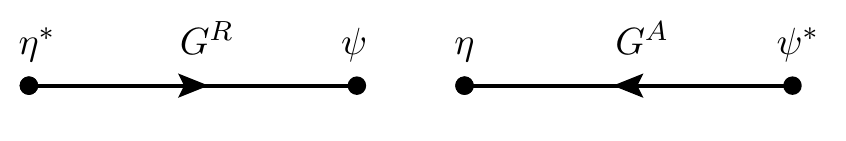}
\caption{The retarded ($G^R$) and advanced ($G^A$) Green's functions. The time arrow runs from left to right.}
\label{fig:blocks0}
\end{figure}

We start with an elementary discussion of the structure of the perturbation theory. We will draw diagrams in such a way that time runs from left to right. Retarded and advanced Green's functions are depicted in Fig.~\ref{fig:blocks0}.

\begin{figure}[rb]
\setlength{\unitlength}{2.3em}
\includegraphics[width=11\unitlength]{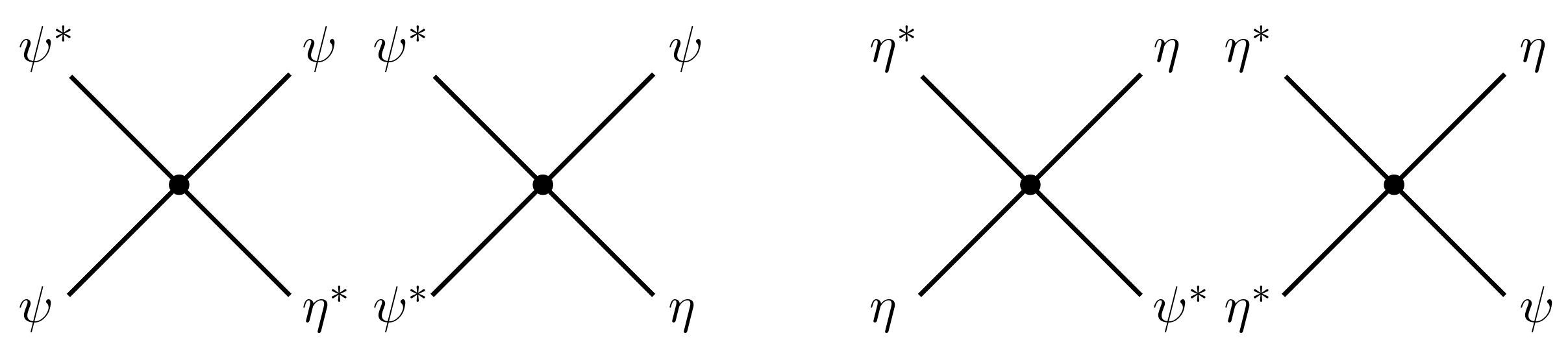}
\caption{In a Keldysh many-body approach to interacting bosons two classes of vertices appear, the classical vertices shown on the left and the quantum vertices shown on the right. In the MSR-type approach used in this paper only the classical vertices are present.}
\label{fig:blocks2}
\end{figure}

\begin{figure}[tb]
\setlength{\unitlength}{2.3em}
\includegraphics[width=10\unitlength]{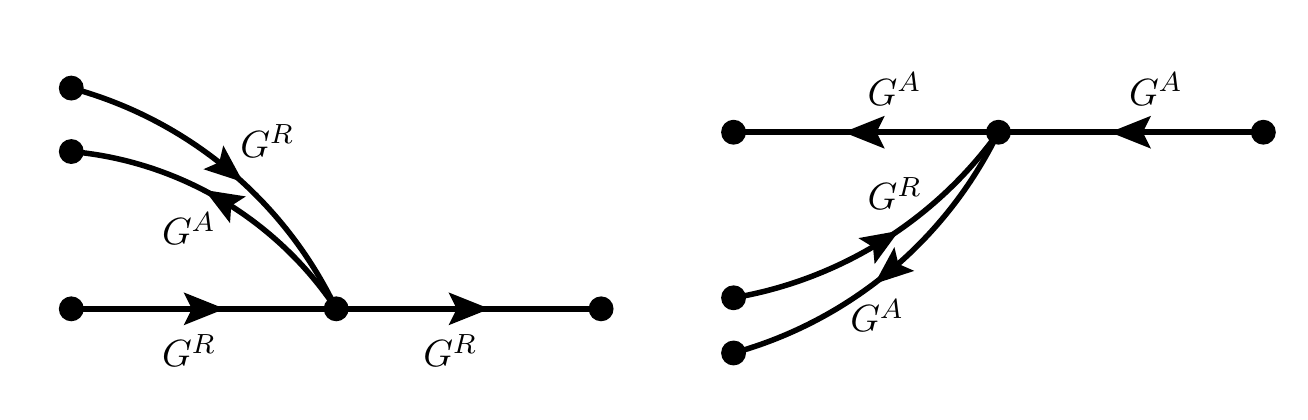}
\caption{Upon averaging with respect to the fields $\psi$ and $\eta$, the (classical) interaction vertices in our approach give rise to the two sub-diagrams shown above.}
\label{fig:blocks1}
\end{figure}

The close similarity to a Keldsyh field theory has already been stressed above. The main difference compared to a full quantum theory of interacting bosons in the Keldysh approach is that out of the two types of vertices depicted in Fig.~\ref{fig:blocks2}, only one is realized. Namely, only the so-called classical vertices, shown on the left hand side of Fig.~\ref{fig:blocks2}, appear in the theory considered here, while the so-called quantum vertices, shown on the right hand side, are absent (see the related discussion in Ref.~\onlinecite{Polkovnikov03}). This has important consequences. It immediately implies that the interaction vertices related to the nonlinearity have the structure shown in Fig.~\ref{fig:blocks1}. This structure, in turn, implies that there are no closed loops in this representation. In order to draw more complex diagrams in a convenient way, we will often depict the interaction vertices with an additional wiggly line (as, for example, in Fig.~\ref{fig:debye} below), but one should keep in mind that the interaction is in fact local in space and instantaneous.

In order to further elucidate the structure of the perturbation theory, we study the expression for the density evolution. The disorder averaging is postponed until the next section, in this section all Green's functions are unaveraged and explicitly depend on the disorder potential. First, we introduce two real Hubbard-Stratonovich fields $\vartheta_{cl}$ and $\vartheta_{q}$, which we assemble into the following matrix:
\be
\hat{\vartheta}=\left(\ba{cc}\vartheta_{cl}&0\\\vartheta_q&\vartheta_{cl}\ea \right).
\ee
With the help of this matrix, the interaction can be represented as
\be
\exp\left(iS_{int}\right)=\left\langle\exp(iS_\vartheta)\right\rangle_\vartheta,
\ee
where we introduced the notation
\be
S_\vartheta=-\int d\bfr dt\;\overline{\phi}(\bfr,t)\hat{\vartheta}(\bfr,t)\phi(\bfr,t),
\ee
and
$\left\langle \dots\right\rangle_\vartheta$ symbolizes the the following averaging procedure
\be
\left\langle \dots\right\rangle_\vartheta=\frac{1}{\mathcal{N}}\int D\vartheta\;\left(\dots\right)\mbox{e}^{\frac{i}{2\lambda}\int d\bfr dt\;\vartheta^T(\bfr,t)\sigma_x\vartheta(\bfr,t)}.\label{eq:vartheta}
\ee
In this equation, $\vartheta=(\vartheta_q,\vartheta_{cl})$ and $\mathcal{N}$ is a normalization constant which we will suppress from now on.

Formula (\ref{eq:vartheta}) implies that fields $\vartheta_q$ and $\vartheta_{cl}$ couple to each other, but not among themselves.
The field $\vartheta_{cl}$ enters $S_\vartheta$ like a classical potential. The quantum component $\vartheta_q$ couples retarded and advanced Green's functions in a specific way.
Taken together, these observation imply that all possible diagrams have the structure indicated in Fig.~\ref{fig:debye}. It is also instructive to further integrate in $\phi$. The result is
\be
&&n(\bfr,t)=\no\\
&&\Big\langle \int d\bfr_1 d\bfr_2\;
\Psi_0^*(\bfr_1)G_{\vartheta_{cl}}^A(\bfr_1,\bfr;0,t)G^R_{\vartheta_{cl}}(\bfr,\bfr_2;t,0)\Psi_0(\bfr_2)\no\\
&&\times\mbox{e}^{ i\int
d\bfr_3d\bfr_4\;\overline{\Psi}_0(\bfr_3)[G^A_{\vartheta_{cl}}\bullet\vartheta_q\bullet
G^R_{\vartheta_{cl}}](\bfr_3,0,\bfr_4,0)\Psi_0(\bfr_4)}\Big\rangle_\vartheta
\ee
We used the triangular structure of $G$ in order to obtain this result. The filled circle $\bullet$ symbolizes a convolution in space and time. The retarded and advanced Green's function in the presence of the classical field $G_{\vartheta_{cl}}$ fulfill the differential equation
\be
&&\left(i\partial_t-\hat{H}-\vartheta_{cl}(\bfr,t)\right)G^{R/A}_{\vartheta_{cl}}(\bfr,\bfr',t,t')\no\\
&&=\delta(\bfr-\bfr')\delta(t-t').
\ee
and $\hat{H}=-\nabla^2/2m+u(\bfr)$. Before averaging in $\vartheta$, the pre-exponential factor describes the evolution of the density on the background of an external classical potential $\vartheta_{cl}$. The exponential contains a similar structure: Each term in the expansion of the exponential symbolizes the evolution of the density up to a certain point. From a formal perspective, integration in $\vartheta_q$ introduces a functional delta function, that fixes $\vartheta_{cl}$ to equal the density.

\begin{figure}[tb]
\setlength{\unitlength}{2.3em}
\includegraphics[width=6.5\unitlength]{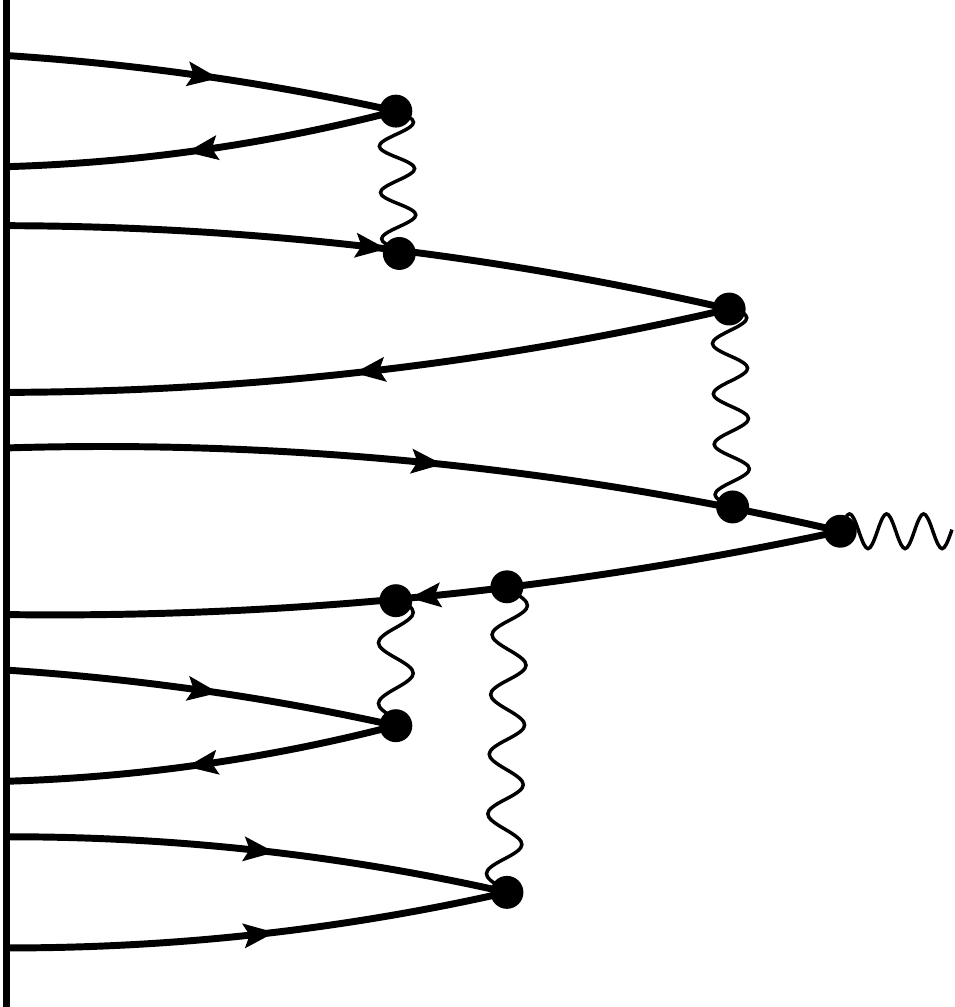}
\caption{General structure of the perturbation theory: The density evolution is represented by the infinite sum of all diagrams of the type displayed in this figure.}
\label{fig:debye}
\end{figure}

\section{Time evolution in a disordered medium}
\label{Sec:timeevolution}
Initially, the disorder potential $u(\bfr)$ enters the action in the form $S_{dis}'=-\int d\bfr dt \;\overline{\phi}(\bfr,t)u(\bfr)\phi(\bfr,t)$.
Disorder averaging with respect to the probability distribution (\ref{eq:distribution}) introduces an effective interaction of the fields
\be
S_{dis}=\frac{i}{2}\gamma\int d\bfr
dt_1dt_2\;\overline{\phi}(\bfr,t_1)\phi(\bfr,t_1)\overline{\phi}(\bfr,t_2)\phi(\bfr,t_2).\no\\\label{eq:sdis}
\ee
This effective interaction is local in space, but non-local in time.

It is usually not possible to take into account disorder effects exactly and one needs to employ approximation schemes. Disorder averaging introduces a quartic term in the action $S$, namely $S_{dis}$ of Eq.~(\ref{eq:sdis}). Here we will treat this term in the self-consistent Born approximation (SCBA), which relies on the weak disorder condition $\tilde{\varepsilon}\tau\gg 1$, where $\tilde{\eps}$ is the characteristic scale for the kinetic energy in the problem.

The SCBA consists in replacing $S_{dis}$ given in Eq.~(\ref{eq:sdis}) by
\be
&&\overline{S}_{dis}\label{eq:overlineSdis}\\
&=&i\gamma\int d\bfr
dt_1dt_2\;\overline{\phi}(\bfr,t_1)\left\langle\phi(\bfr,t_1)\overline{\phi}(\bfr,t_2)\right\rangle\phi(\bfr,t_2).\no
\ee
The average can be taken in two equivalent ways, which explains the additional factor of $2$ compared to Eq.~(\ref{eq:sdis}). Averaging is performed with respect to the action $S$ \emph{after} the disorder averaging, i.e., self-cosistently. This implies that, generally speaking, the disorder part of the self-energy also implicitly depends on the interaction (namely via the Green's function $-i\left\langle\phi\overline{\phi}\right\rangle$).

\subsection{Noninteracting theory}
\label{sec:noninteracting}
This section contains an elementary discussion of the theory for the density evolution in the noninteracting case $\lambda=0$. It serves as a preparation for the discussion of the interacting model. Furthermore, we use the opportunity to introduce our notation and to stress the most important differences to the calculation of the density-density correlation function in disordered electron systems.

In this case one obtains
\be
&&n(\bfr,t)=\int d\bfr_1 d\bfr_2
\left\langle \Psi_0^*(\bfr_1)G_{0}^A(\bfr_1,\bfr;0,t)\right.\label{eq:dens}\\
&&\left.G^R_{0}(\bfr,\bfr_2;t,0)\Psi_0(\bfr_2)\right\rangle_{dis}\no
\ee
where $\left(i\partial_t-\hat{H}\right)G^{R/A}_{0}(\bfr,\bfr',t,t')=\delta(\bfr-\bfr')\delta(t-t')$.

In the SCBA, the disorder averaged Green's function is given by
\be
\underline{G}^{R/A}(\bfp,\eps)=\left(\eps-\frac{p^2}{2m}\pm \frac{i}{2\tau(\eps)}\right)^{-1},\label{eq:disorderedG}
\ee
where
\be
\tau_\eps^{-1}=2\pi\nu(\eps)\gamma \label{eq:taueps}
\ee
is the scattering rate and $\nu(\eps)$ is the density of states. This result is obtained as follows. For $\lambda=0$, the defining relation for the disorder part of the self-energy in the SCBA is
\be
\Sigma^{R/A}_{dis}(\eps)=\gamma \int (d\bfp) \frac{1}{\eps-\eps_\bfp-\Sigma_{dis}^{R/A}(\eps)}
\ee
Here and in the following we use the notation $(d\bfp)=d^d p/(2\pi)^d$. The scattering time $\tau_\eps$ is defined as
\be
\Im[\Sigma_{dis}^{R/A}(\eps)]=\mp1/(2\tau_\eps)\label{eq:imSigma}
\ee
Upon introducing the variable $\xi_\bfp=\eps_\bfp-\eps$ the integration measure transforms as  $\int (d\bfp) =\int_{-\eps}^\infty d\xi_\bfp \nu(\eps+\xi_\bfp)$, where the trivial angular averaging has already been performed. Focusing on the imaginary part of the self-energy first, one may extend the lower limit of the integration in $\xi_\bfp$ to $-\infty$ in the weak disorder limit, $\eps-\Re[\Sigma^R_{dis}(\eps)]\gg1/\tau_\eps$. At the same time, this step regularizes the integral for the real part of the self-energy. The integrand for the imaginary part of the self-energy is strongly peaked around $\xi_\bfp= 0$ and one may  replace $\nu(\eps+\xi_\bfp)\approx \nu(\eps)$ and take the density of states out of the integral. The remaining integral is easily performed and the result is
\be
\Im[\Sigma_{dis}^{R/A}(\eps)]=\mp\pi\nu(\eps) \gamma
\ee
in agreement with (\ref{eq:taueps}) and (\ref{eq:imSigma}).

\begin{figure}
\setlength{\unitlength}{2.3em}
\includegraphics[width=5.5\unitlength]{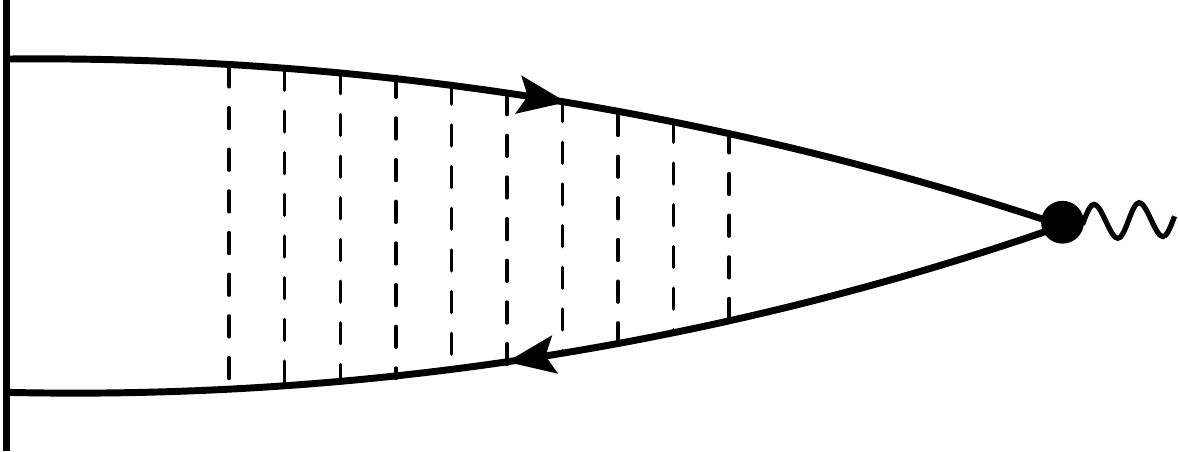}
\caption{Diagrammatic representation of the diffusion process in the absence of the nonlinearity.}
\label{fig:blocks3}
\end{figure}

As is well known \cite{AGD63}, in the leading approximation in $1/\eps\tau$, one should not only replace $G_0$ by $\underline{G}$ in formula (\ref{eq:dens}) for the density, but sum the whole set of diagrams with non-crossing impurity lines as shown in Fig.~\ref{fig:blocks3}. Effectively, this amounts to summing a geometric series. This procedure leads to the expression
\be
&&n(\bfq,\omega)=\int (d\bfp)(d\eps)\;\Psi_0(\bfp_+)\Psi_0^*(\bfp_-)\;\no\\
&&\times \underline{G}^R(\bfp_+,\eps_+)
\underline{G}^A(\bfp_-,\eps_-)\sum_{n=0}^{\infty} L_\eps^n(\bfq,\omega)
\ee
where
\be
L_\eps(\bfq,\omega)&=&\gamma \int (d\bfp_1)
\;\underline{G}^R(\bfp_{1+},\eps_+)\underline{G}^A(\bfp_{1-},\eps_-)\no
\ee
We use the notation $(d\eps)=d\eps/(2\pi)$ for frequency integrals, $\bfp_\pm=\bfp\pm\bfq/2$ and $\eps_\pm=\eps\pm\omega/2$.
The expression is quite similar to the familiar density-density correlation function in electronic systems. Note, however,
that in the latter case the frequency integration is restricted to a small interval around the Fermi surface of order of the temperature by the presence of a distribution function. In contrast, here the {\it momentum} integration is restricted by the initial wave functions $\Psi_0$ and $\Psi^*_0$. The frequency integration, on the other hand, is a priori not limited.

Let us assume that the inequalities
$\varepsilon\tau_\eps\gg 1$, $\omega\tau_\eps\ll 1$, $q l_\varepsilon\ll 1$ are fulfilled (diffusion approximation),
where $l_{\varepsilon}=v_\varepsilon\tau_\eps$ is the mean free path, $v_\varepsilon=p_\varepsilon/m$
and $p_\varepsilon=\sqrt{2m\varepsilon}$ are the velocity and the momentum at energy $\varepsilon$.
In this case we can calculate the sum approximately by using the expansion
\be
L_\eps(\bfq,\omega)\approx
1+i\omega\tau_\eps-l_\varepsilon^2 q^2/2.
\ee
It will be useful to introduce a frequency dependent diffusion
constant as $D_{\eps}=v_\varepsilon^2\tau_\eps/d$ in dimension $d$. After performing the sum in the equation for the density we obtain
\be
n(\bfq,\omega)&=&\int (d\bfp)(d\eps)\;\Psi_0(\bfp_+)\overline{\Psi}_0(\bfp_-)\\
&&\times \underline{G}^R(\bfp_+,\eps_+)\underline{G}^A(\bfp_-,\eps_-)\frac{1}{\tau_\eps}\mathcal{D}_{\eps}(\bfq,\omega),\no
\ee
where the energy dependent diffuson is
\be
\mathcal{D}_{\eps}(\bfq,\omega)=(D_{\eps}\bfq^2-i\omega)^{-1}.
\ee

The next step is to integrate in $\eps$, where one encounters the following integral
\be
\int
(d\eps)
\frac{1}{\eps_+-\eps_{\bfp_+}+\frac{i}{2\tau_\eps}}\;\frac{1}{\eps_--\eps_{\bfp_-}-\frac{i}{2\tau_\eps}}\;\frac{1}{\tau_\eps}\mathcal{D}_{\eps}(\bfq,\omega).\no\\
\label{eq:frequencyintegration}
\ee
For $\eps_{\bfp_+}\sim\eps_{\bfp_-}\gg 1/\tau$ the
most important $\eps$ are of the order of $\eps_p$ and we can perform the
integral with the help of the residue theorem considering the poles originating from the Green's
functions only, thereby effectively replacing $D_{\eps}$ by
$D_{\eps_\bfp}$. A distinction between $\eps_{\bfp_+}$ and
$\eps_{\bfp_-}$ in the argument of the diffusion coefficient would be beyond the accuracy of our approach. The result is
\be
&&n(\bfq,\omega)\approx \int (d\bfp)
F(\bfp,\bfq)\mathcal{D}_{\eps_{\bfp}}(\bfq,\omega),
\label{eq:densofq}
\ee
where we introduced the notation
\be
F(\bfp,\bfq)=\Psi_0(\bfp+\bfq/2)\Psi_0^*(\bfp-\bfq/2).\label{eq:Wigner}
\ee
It is clear from the previous arguments that the approach is valid as long
as $\eps_\bfp\tau_{\eps_\bfp} \gg 1$. Typical momenta $\bfp$ are controlled by the initial wave-function $\Psi$. For the averaged density as a function of coordinates and time we find the expression
\be
&&n({\bf r},t)=\label{eq:nxx}\\
&&\int (d\bfp)\;\frac{\Theta(t)}{4\pi D_{\varepsilon_\bfp} t} \int d{\bf r}_1\;\mbox{e}^{-({\bf r}-{\bf r}_1)^2/(4D_{\varepsilon_\bfp} t)} F(\bfp,{\bf r}_1).\no
\ee
For $|\bfr_1|\ll |\bfr|$, i.e. for distances $|\bfr|$ exceeding by far the extension of the initial wave-packet, we may neglect $\bfr_1$ in the exponent and obtain
\be
n(\bfr,t)=\Theta(t)\int
(d\bfp)\frac{|\Psi_0(\bfp)|^2}{4\pi t
D_{\varepsilon_\bfp}}\;\mbox{e}^{-\bfr^2/(4tD_{\varepsilon_\bfp})}.
\label{eq:Shapiro}
\ee
This expression was presented in Ref. \onlinecite{Shapiro07}.

In the calculation described in this section, the frequency-integration was performed \emph{before} the momentum integration in $\bfp$ (see Eq. \ref{eq:frequencyintegration}). Relevant momenta in the integral of Eq. (\ref{eq:densofq}) are determined by $F(\bfp,\bfq)$, which encodes the information contained in the initial wave-function $\Psi$. For a generalization to the interacting case, it will be more useful to perform the integration in $\bfp$ before the integration in $\eps$. In order to achieve this goal, we introduce the distribution function $f$ in the following way
\be
f(\bfr,t_1,t_2)&=&\gamma\int d\bfr_3d\bfr_4 \;\underline{G}^R(\bfr_1-\bfr_3,t_1)\no\\
&&\times\Psi(\bfr_3)\Psi^*(\bfr_4)\underline{G}^A(\bfr_4-\bfr_1,-t_2).\quad\label{eq:distr}
\ee
It describes the initial section of the diffusion ladder, compare Fig.~\ref{fig:blocks3}.
With the help of this definition one can write
\be
n(\bfr,t)=\int(d\eps)n(\bfr,\eps,t),
\ee
where the energy resolved density is
\be
n(\bfr,\eps,t)=2\pi\nu(\eps)\int_{\bfr_1} \mathcal{D}_{\eps}(\bfr-\bfr_1,t-t_1)f(\bfr_1,\eps,t)\;\label{eq:nxy}
\ee
and
\be
f(\bfr,\eps,t)=\int d(\Delta t) f(\bfr,t+\Delta t/2,t-\Delta t/2)\;\mbox{e}^{i\eps\Delta t}.\quad\label{eq:feps}
\ee
We can make contact with the previous results of this section by noting that for times $t\gg \tau$ one can approximate (see Appendix \ref{app:distribution})
\be
2\pi\nu(\eps) f(\bfr,\eps,t)\approx \delta(t)F(\eps,\bfr),\label{eq:distrappr}
\ee
where
\be
F(\eps,\bfr)=\int(d \bfp)\;F(\bfp,\bfr)\;(2\pi)\delta(\eps-\eps_{\bfp}).\label{eq:Fxx}
\ee
This concludes our discussion of the non-interacting theory.

As for the electronic systems, it is most convenient to formulate the microscopic theory with the help of a frequency dependent distribution function, since momentum is not conserved during the scattering process. At the same time, the initial distribution is determined by the momentum dependence of the wave function, Eq.~(\ref{eq:Wigner}). In the quasiparticle approximation, one can translate between the two representations, Eq.~(\ref{eq:Fxx}). The specifics of the given problem in comparison with diffusion in a degenerate electronic system is that the dependence of the diffusion coefficient needs to be kept explicitly. Each particle at a given energy diffuses with its own diffusion coefficient and the total density is obtained through a convolution with the distribution function, Eqs.~(\ref{eq:nxx}) and (\ref{eq:nxy}). The fact that the energy distribution may be broad has the important consequence that the density may differ considerably from the form $n(\bfr,t)\propto\exp(-c\bfr^2/t)$ (with a constant $c$), which holds for diffusion at a fixed energy. To illustrate this important point, we briefly discuss an example first introduced in Ref.~\onlinecite{Shapiro07}.

\subsubsection{Gaussian initial distribution}

As an instructive example, one can easily calculate the asymptotic distribution for the initial condition \cite{Shapiro07}
\be
|\Psi_0(\bfp)|^2=(2\pi)^2\frac{N}{\pi}\frac{1}{k_0^2}\mbox{e}^{-\bfp^2/k_0^2}
\ee
It is convenient to introduce a typical diffusion coefficient
$D_0=D_{k_0}$. One may use Eq.~(\ref{eq:Shapiro}) to find\cite{Shapiro07}
\be
n(\bfr,t)=\Theta(t)\frac{N}{2\pi D_0
t}\;K_0\left(r/\sqrt{D_0t}\right),
\ee
which decays asymptotically as $n(\bfr,t)\propto\exp(-r/\sqrt{D_0t})$ for $r\gg\sqrt{D_0t}$.
This should be compared to the case where the diffusion coefficient $D_0$ is momentum-independent and one finds (in $2d$) $n(\bfr,t)\propto \exp\left(-r^2/4D_0t\right)$. We see, that the asymptotic profile depends crucially on the initial distribution of momenta. Consequently, a detailed knowledge of initial conditions is required for the interpretation of experiments.

\subsection{Diagrammatic perturbation theory for the nonlinear problem and the kinetic equation}
\label{sec:intperturbation}

One can organize a systematic perturbation theory for the nonlinear problem ($\lambda\ne 0$) in the limit of weak disorder. This regime is characterized by the condition $\eps\tau\gg 1$, where $\eps$ is the characteristic energy determining the diffusion coefficient. In this paper, we make use of the fact that in two spatial dimensions and for   weak disorder one expects an extended regime for which the density evolution is diffusive, i.e. we are interested in nonlinear \emph{diffusion} and do not consider localization effects. We may therefore restrict ourselves to the leading order in the smallness paramter $1/(\eps\tau)$. At this level of accuracy, the standard diagrammatic technique can be used to select diagrams for which impurity lines do not cross.

In contrast to the noninteracting case discussed in the previous section, for which a single diffusion mode was sufficient for the description, the nonlinearity introduces an effective coupling of diffusion modes to each other. This coupling is not completely arbitrary, but must be consistent with the conservation of the total density in the limit of vanishing momentum.

\begin{figure}[h]
\setlength{\unitlength}{2.3em}
\includegraphics[width=8\unitlength]{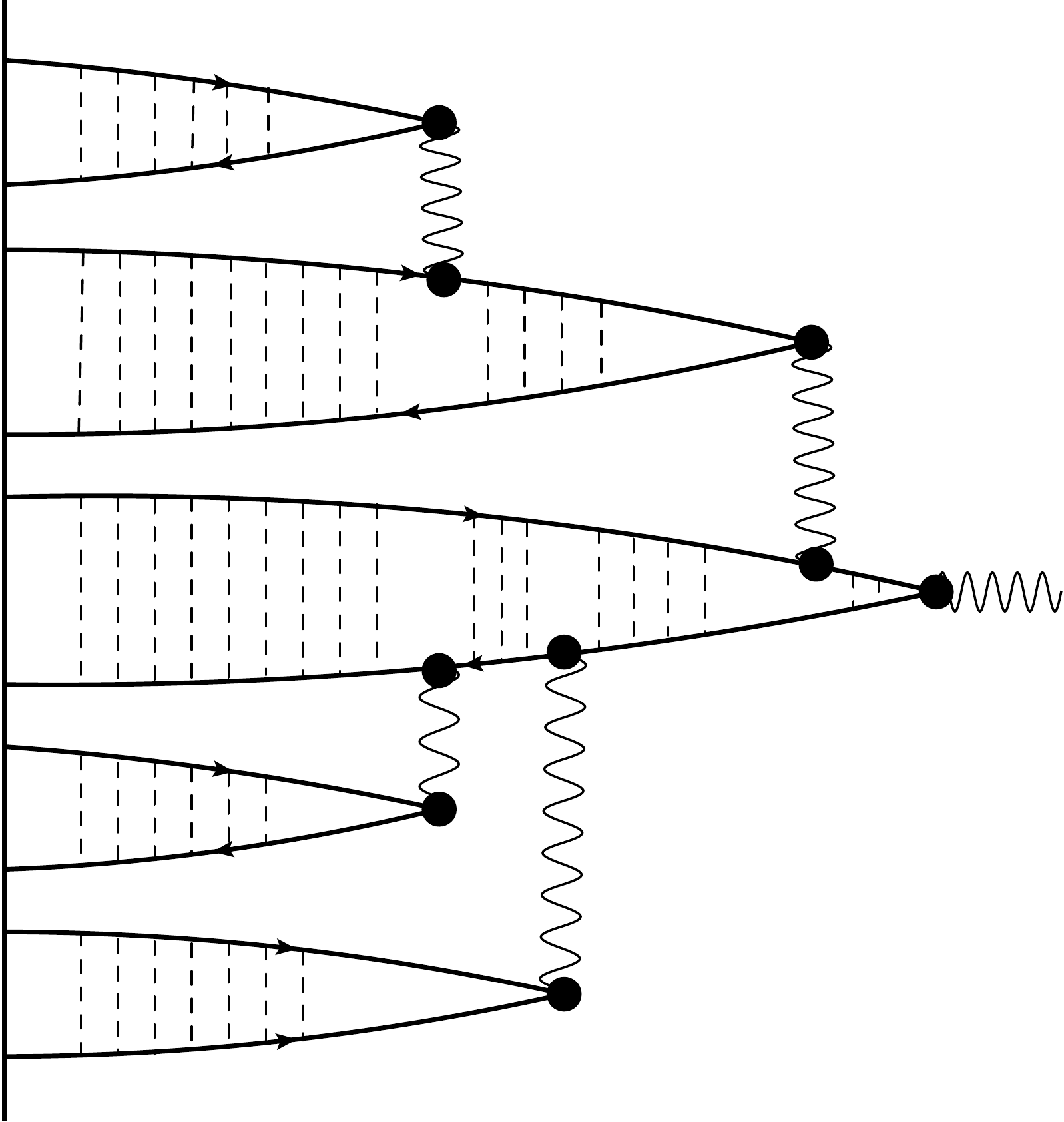}
\caption{On a diagrammatic level, the solution of the kinetic equation corresponds to the sum of all diagrams of the type shown in this figure.}
\label{fig:Cascade}
\end{figure}

The relevant diagrams of perturbation theory are of the form depicted in Fig.~\ref{fig:Cascade}, where the left hand side is associated with the initial  distribution function and requires a separate consideration, see Appendix \ref{app:distribution}. Each skeleton diagram, by which we mean a diagram of the form shown in Fig. \ref{fig:debye}, i.e., before disorder averaging, can be dressed by disorder in several equivalent ways, namely, each vertex is associated with a combinatorial factor 2. This is related to the fact that the interaction is chosen to be local in space [although we draw extended interaction lines in order to have a more convenient graphical representation]. This combinatorial factor is taken care of by choosing the decoupling in Eq. \ref{eq:slow} below in two equivalent ways.

We see that the expansion takes the form of a self-consistent Hartree theory. Due the self-consistency, the structure of the theory reveals itself already at the first order of perturbation theory in $\lambda$.

In the following we will discuss the first order perturbation theory and explain the origin of the different terms in the kinetic equation on this level. To this end, consider the diagrams in Fig. \ref{fig:Cascade}. The interaction line can couple both to the retarded and the advanced Green's function and due to important cancellations among these two the diagrams should always be grouped in pairs. The interaction line carries both momenta and frequencies and these can be considered as small, since they are related to the adjoint diffuson, or, in more physical terms, since the density is smooth and slowly varying in time. For pedagogical reasons, we will separate the discussion into two parts, the transfer of small momenta at vanishing frequency and that of small frequencies for vanishing momentum transfer.

We start with a finite momentum transfer. Here, the important point is that the diffuson to the left depends on the relative momentum $\bfq$ of the retarded and advanced Green's function only, but not on the sum of momenta. This relative momentum $\bfq$ is the same irrespective of whether the interaction line goes to the retarded or the advanced Green's function.

As far as frequencies are concerned,  the diffuson to the left of the block depends not only on the relative frequency, but also -- via the diffusion coefficient -- on the center of mass frequency. Therefore, it distinguishes between the two diagrams.

\begin{figure}[h]
\setlength{\unitlength}{2.3em}
\includegraphics[width=8\unitlength]{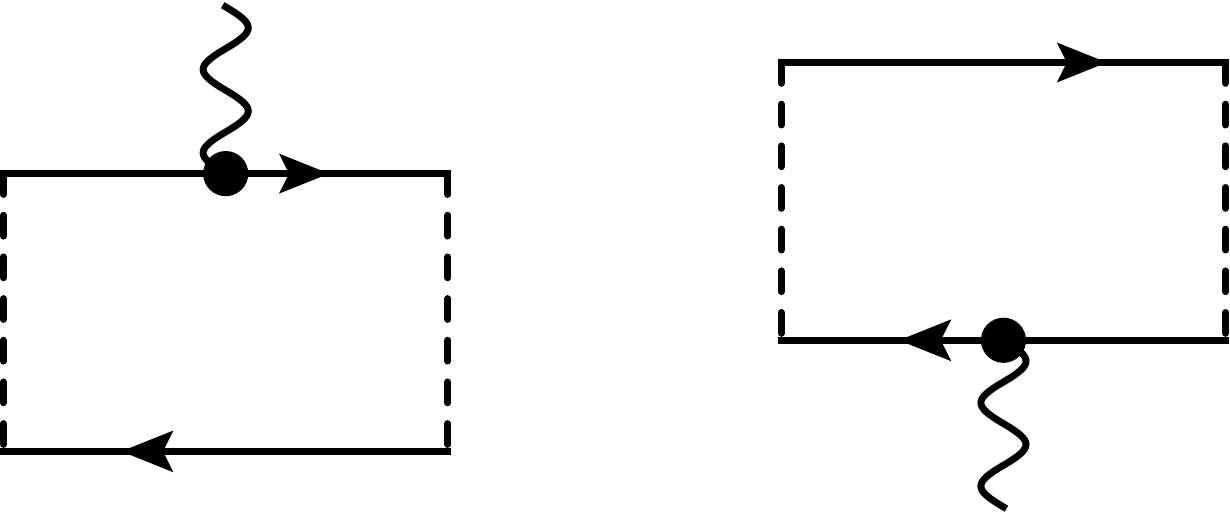}
\caption{The two box diagrams, $\mathcal{B}_R$ to the left and $\mathcal{B}_A$ to the right. The interaction line carries frequency $\omega_1$ and momentum $\bfq_1$ (incoming). For the individual diagrams a constant term remains in the limit of vanishing external frequencies and momenta. This constant cancels, however, between the two diagrams. The cancellation is related to number conservation as is discussed in the main text.}
\label{fig:BRBA}
\end{figure}

Let us introduce the expressions for the box in the two cases (see also Fig. \ref{fig:BRBA})
\be
\mathcal{B}_R(\bfq,\bfq_1,\omega,\omega_1)&=&\frac{1}{2\pi\nu \tau^2}\int(d\bfp)\underline{G}^R(\bfp_+-\bfq_1,\eps_+-\omega_1)\no\\
&&\times G^A(\bfp_-,\eps_-)G^R(\bfp_+,\eps_+)
\ee
and $\mathcal{B}_A(\bfq,\bfq_1,\omega,\omega_1)=\mathcal{B}^*_R(-\bfq,-\bfq_1,-\omega,-\omega_1)$. The dependence on the spectator argument $\eps$ will be suppressed.

Then the energy resolved densities for the two diagrams read
\be
&&n_{1R/L,\eps}(\bfq,\omega)=\no\\
&&\quad\mathcal{D}_{\eps}(\bfq,\omega)\int (d\bfq_1)(d\omega_1)\;n_{0,\eps\mp\frac{\omega_1}{2}}(\bfq-\bfq_1,\omega-\omega_1)\no\\
&&\quad\times \vartheta(\bfq_1,\omega_1)\mathcal{B}_{R/L}(\bfq,\omega,\omega_1)
\ee
where we denoted the noninteracting energy resolved density (compare Sec.~\ref{sec:noninteracting}) as
\be
n_{0,\eps}(\bfq,\omega)=2\pi\nu f_{\eps}(\bfq,\omega)\mathcal{D}_{\eps}(\bfq,\omega)
\ee
and also used its relation to the density $n_0(\bfr,t)=\int (d\eps)n_{0,\eps}(\bfr,t)$ in the linear case when introducing the notation
\be
\vartheta(\bfr,t)=2\lambda n_{0}(\bfr,t)
\ee
As will become clear in the following, $\vartheta(\bfr,t)$ can be interpreted as an effective potential.

The averaged density at order $\lambda$  is the sum of the two densities $n_1=n_{1L}+n_{1R}$. It is
\be
&&n_{1}(\bfq,\omega)=\mathcal{D}_{\eps}(\bfq,\omega)\int(d\bfq_1)(d\omega_1)\vartheta(\bfq_1,\omega_1)\\
&&\times\Big[n_{0\eps}(\bfq-\bfq_1,\omega-\omega_1)[\mathcal{B}_R+\mathcal{B}_L](\bfq,\bfq_1,\omega,\omega_1)\no\\
&&\quad-\omega_1\partial_{\eps}n_{0\eps}(\bfq-\bfq_1,\omega-\omega_1)\frac{1}{2}[\mathcal{B}_R-\mathcal{B}_L](\bfq,\bfq_1,\omega,\omega_1)\Big]\no
\label{eq:firstorder}
\ee
By explicit calculation one finds in the limit of small momenta and frequencies
\be
&&[\mathcal{B}_R+\mathcal{B}_L](\bfq,\bfq_1,\omega,\omega_1)\approx\frac{\tau}{m}\bfq(\bfq-\bfq_1)\\
&&[\mathcal{B}_R-\mathcal{B}_L](\bfq,\bfq_1,\omega,\omega_1)\approx-2i
\ee

Let us start the discussion with the case of finite momentum transfer.
Here, the combination $\mathcal{B}_R+\mathcal{B}_A$ enters the diagram and one immediately finds that the leading constant term cancels and the coupling is proportional to $\bfq(\bfq-\bfq_1)$. In particular, it is proportional to the external momentum $\bfq$. The cancellation of the constant term is not accidental, but enforced by number conservation. Indeed, the limit $\bfq\rightarrow 0$ is related to the conservation law for the total density. This can be seen best in the language of the kinetic equation discussed below. In fact, it turns out that the combination $\mathcal{B}_R+\mathcal{B}_A$ still contains a small constant term of order $1/(\tilde{\eps}\tau)^2$, which disappears, however, when one uses the full $\vartheta$ dependent Green's function for the self-consistent Born approximation as is automatically the case in the kinetic equation approach described in Sec.~\ref{sec:quasiclassics}.

Turning to the finite frequency transfer next, we see that the situation is different. Here, the constant of the box diagrams $\mathcal{B}_R$ and $\mathcal{B}_L$ may contribute and the result is proportional to the difference of diffusons with different center of mass energies in agreement with our previous discussion.

Proceeding towards the kinetic equation next, one may multiply Eq.~\ref{eq:firstorder} by $\mathcal{D}^{-1}_{\eps}(\bfq,\omega)$ and we present it together with the real space representation of the equation for $n_{0,\eps}$
\be
(\partial_t-D_\eps\nabla^2)n_{1,\eps}&=&-[\frac{\tau}{m}(\nabla\vartheta\nabla)n_{0\eps}+\partial_t\vartheta\partial_\eps n_{0\eps}]\no\\
(\partial_t-D_{\eps}\nabla^2)n_{0,\eps}&=&2\pi\nu f(\bfr,t,\eps)
\ee
We easily recognize the first iterative solution to the kinetic equation, once we use the relation between $f$ and $F$ discussed in Appendix \ref{app:distribution}. We will not follow this route further and formally sum up all diagrams, although this can be done.  It has become clear that an equation is much more useful then any finite order in perturbation theory and there are more effective ways to derive the kinetic equation.

\subsection{Slow mode decomposition}
\label{Sec:slow}

As a first step in deriving the kinetic equation we turn to the interaction term $S_{int}$ specified in Eq.~(\ref{eq:sint}). The self-consistent potential $\vartheta(\bfr,t)=2\lambda n(\bfr,t)$ is introduced in the following way. We average $S_{int}$ and obtain
\be
&&\overline{S}_{int}\label{eq:overlineSint} \label{eq:slow}\\
&=&-2\lambda\int d\bfr dt \;\overline{\phi}(\bfr,t)\left\langle[\phi(\bfr,t)\;\overline{\phi}(\bfr,t)]_{21}\right\rangle\phi(\bfr,t)\no\\
&=&-\int d\bfr dt \;\overline{\phi}(\bfr,t)\vartheta(\bfr,t)\phi(\bfr,t),\no
\ee
where $n(\bfr,t)=\left\langle[\phi(\bfr,t)\;\overline{\phi}(\bfr,t)]_{12}\right\rangle$ was used in the last step. The averaging $\left\langle\dots\right\rangle$ in both Eqs. (\ref{eq:overlineSdis}) and (\ref{eq:overlineSint}) is defined self-consistently, namely with respect to $\overline{S}\equiv S_{s}+S_{0}+\overline{S}_{dis}+\overline{S}_{int}$. Let us stress that this approach includes interaction effects non-perturbatively as a result of self-consistency. In comparison with the clean case an additional factor of $2$ appears in the definition of the self-consistent field $\vartheta$. This is not a double counting, but a result of a typical slow-mode decomposition, in this case in the density channel. Indeed, it will be valid only if $\vartheta$ is a slowly varying field, it means that momenta of the fields $\phi$ and $\overline{\phi}$ are close to each other. In principle, one could also consider  "anomalous" averages of the type $\left\langle \psi(\bfr,t)\psi(\bfr,t)\right\rangle$ and $\left\langle \psi^*(\bfr,t) \psi^*(\bfr,t)\right\rangle$. For systems for which the potential energy is not much smaller than the kinetic energy, such averages can in principle become important. In the limit we consider, namely for $\overline{\eps}\gg \lambda n$, these terms are, however, less effective than the potential $\vartheta$ as already argued in Sec.~{\ref{sec:intro}}.

\subsection{Green's function}
After treating both disorder and interaction self-consistently as described in the previous section we obtained the action $\overline{S}$. Due to the presence of the source terms describing the injection process, the fields $\psi$ and $\psi^*$ have non-vanishing expectation values. This inconvenient feature can easily be cured by shifting the fields appropriately. To this end, we introduce the Green's function $G$ as the average $G=-i\left\langle\phi\overline{\phi}\right\rangle_{\tilde{S}}$, where the averaging is with respect to $\tilde{S}=\overline{S}-S_{s}$. This immediately implies $\tilde{S}=\int\;\overline{\phi}\;G^{-1}\phi$. We can define $G$ explicitly by writing its inverse
\be
G^{-1}(\bfr_1,t_1,\bfr_2,t_2)&=&g_0^{-1}(\bfr_1-\bfr_2,t_1-t_2)\label{eq:G1}\\
&+&i\gamma \left\langle \phi(\bfr_1,t_1)\overline{\phi}(\bfr_2,t_2) \right\rangle_{\overline{S}}\no\\
&-&\vartheta(\bfr_1,t_1) \;\delta(\bfr_1-\bfr_2)\delta(t_1-t_2).\no
\ee
By denoting the averaging with the label $S$ in this equation, we want to remind that it should be performed with respect to $\overline{S}$, not $\tilde{S}$. After introducing the shifted fields
\be
\zeta(\bfr_2,t_2)&=&\phi(\bfr_2,t_2)-i\int d\bfr_3\;G(\bfr_2,t_2,\bfr_3,0)\phi_0(\bfr_3)\no\\
\overline{\zeta}(\bfr_1,t_1)&=&\overline{\phi}(\bfr_1,t_1)+i\int d\bfr_3 \;\overline{\phi}_0(\bfr_3)G(\bfr_3,0,\bfr_1,t_1),\no\\
\ee
we observe that $\overline{S}=\int \overline{\zeta}\;G^{-1}\zeta$, i.e. $G=-i\left\langle \zeta\overline{\zeta}\right\rangle_{\overline{S}}$. We used the fact that $G_{21}=0$ when completing the square. Let us also note that
\be
\left\langle \phi \overline{\phi} \right\rangle_{\overline{S}}=\left\langle \zeta\overline{\zeta}\right\rangle_{\overline{S}}+\sigma_{+}\frac{1}{\gamma}F=iG+\sigma_+\frac{1}{\gamma} F,\label{eq:shift1}
\ee
where
\be
F(\bfr_1,\bfr_2,t_1,t_2)&=&\gamma\int d\bfr_3d\bfr_4 \;G_{R}(\bfr_1,t_1,\bfr_3,0)\no\\
&&F_0(\bfr_3,\bfr_4)G_{A}(\bfr_4,0,\bfr_2,t_2)\label{eq:F}
\ee
and $F_0(\bfr_3,\bfr_4)=\Psi_0(\bfr_3)\Psi^*_0(\bfr_4)$. In particular
\be
n(\bfr,t)=iG^{12}({\bf r},t,{\bf r},t)+\frac{1}{\gamma}f(\bfr,t,t),\label{eq:nfromG}
\ee
where we denoted
\be
f(\bfr,t_1,t_2)=F(\bfr,\bfr,t_1,t_2).\label{eq:f}
\ee
By inserting relation (\ref{eq:shift1}) into (\ref{eq:G1}) we obtain an equation for $G$ in the form
\be
&&\left(i\partial_{t_1}-\hat{\varepsilon}_{1}-\vartheta({\bf r}_1,t_1)\right)G({\bf r}_1,t_1,{\bf r}_2,t_2)\no\\
&&-\int dt_3 \Sigma({\bf r}_1,t_1,t_3) G({\bf r}_1,t_3,{\bf r}_2,t_2)\no\\
&&=\delta({\bf r}_1-{\bf r}_2)\delta(t_1-t_2)\label{eq:greens}.
\ee
where $\hat{\varepsilon}_1$ is the operator of the kinetic energy acting on coordinate $\bfr_1$.
Let us comment on the different terms entering the equation. The $11$ and $22$ components of the matrix $G$ are retarded and advanced Green's functions, respectively, for which we use the notation $G^R$ and $G^A$. Disorder effects are included within the framework of the self-consistent Born approximation which gives rise to a contribution to the self-energy,
\be
\Sigma_{dis}({\bf r},t_1,t_2)=\gamma G({\bf r},t_1,{\bf r},t_2),
\ee
The total self-energy
\be
\Sigma=\Sigma_{dis}+\Sigma_{s}
\ee
additionally comprises a source term $\Sigma_s$, which is purely off-diagonal and related to the initial conditions. It can be written as
\be
\Sigma_{s}({\bf r},t_1,t_2)=-i\sigma_{+}f({\bf r},t_1,t_2),
\ee
where $\sigma_+=(\sigma_x+i\sigma_y)/2$, and $f$ is defined through Eqs.~(\ref{eq:f}) and (\ref{eq:F}).

The equation for the Green's function (\ref{eq:greens}) is fully consistent with Eq.~(\ref{eq:disorderedG}) for the noninteracting case. Here, however, $G^{R/A}$ depend on the classical self-consistent potential. Besides, the dependence on the initial conditions is explicitly included in the definition. The function $f$ plays the role of the initial distribution function in our description. The density is expressed in terms of the components of $G$ as  shown in Eq.~(\ref{eq:nfromG}).

Thus we arrive at two equations for $G$ and $n$, that are coupled by the self-consistency relation $\vartheta=2\lambda n$. The first term in the Eq. (\ref{eq:nfromG}) for $n({\bf r},t)$ accounts for diffusion for times much larger than  $t\gg \tau$, while the second term is a short range contribution that describes the initial
expansion up to times of the order of the scattering time $\tau$. It turned out to be possible to organize both the differential equation and the relation between the density $n$ and the components of $G$ in such a way that the information about the initial wave function always appears together with $G^R$ and $G^A$. Recall that $G^R$ and $G^A$ are separately averaged over disorder.

The equation for the Green function, Eq.~(\ref{eq:greens}), still contains more information than is needed for calculating the density evolution and hence further
simplifications can be made. In essence, we will proceed in analogy to the quasi-classical approximation widely used in the theory of nonhomogeneous superconductivity \cite{Eilenberger68,Larkin68,Kopnin01}.

\subsection{Quasiclassical approximation}

As is well known, for the analysis of the effects of weak
disorder, $1/\tau \ll \varepsilon $ and for smooth external perturbations (on the scale of wave length), one may pass from the full quantum mechanical
equations to a reduced quasiclassical description. In the case of superconductivity, this procedure leads from the Gor'kov equations to the
Eilenberger equation in the ballistic limit and, further on, to the Usadel equation in the diffusive limit. Following this route, we will derive
an Usadel-like diffusive equation for a wave-packet evolving in the self-consistent potential which arises as a result of the nonlinearity.  The obtained kinetic equation determines the distribution function $n(\bfr,t,\eps)$, from which the density of the gas at a given moment and spatial coordinate is found as $n(r,t)=\int (d\eps) \;n(\eps,\bfr,t)$.

\label{sec:quasiclassics}

We start by introducing a mixed (Wigner) representation for the Green's function, \be
G({\bf r}_1,{\bf r}_2,t_1,t_2)&=&\int (d\bfp)(d\eps)\;G({\bf r},{\bf p},t,\varepsilon)\no\\
&&\quad\times\mbox{e}^{i{\bf p}({\bf r}_1-{\bf r}_2)-i\varepsilon(t_1-t_2)},
\ee
where ${\bf r}=({\bf r}_1+{\bf r}_2)/2$ and $t=(t_1+t_2)/2$. Considering first the linear case, $\lambda=0$, the frequency defines a momentum scale
$p_\varepsilon=\sqrt{2m\varepsilon}$, wavelength $\lambda_\varepsilon=2\pi/p_\varepsilon$ and time scale $t_\varepsilon=\varepsilon^{-1}$.
Initially, the typical scale for $\varepsilon$ is determined by the function $f({\bf r},t,\varepsilon)$, which in turn reflects the momentum distribution of the injected wave-packet, see Eq.~(\ref{Eq:Distr}) below. If the density and self-energies are smooth on the scale $\lambda_\varepsilon$, the Green's function can be averaged on this scale. A necessary prerequisite is that $\varepsilon$ is sufficiently large. In this sense, $\varepsilon$ plays a role similar to the Fermi-energy in electronic systems. In the same spirit, the weak disorder condition, which is needed to formally justify the use of the self-consistent Born approximation, can be formulated as $\varepsilon\tau\gg 1$. The averaging alluded to above can be implemented by integrating the Green's function in deviations from $p_\varepsilon$. In the nonlinear case $\lambda\ne 0$, the frequency $\varepsilon$ in the previous argument should be replaced by
\be
\tilde{\varepsilon}({\bf r},t)=\epsilon-\vartheta({\bf r},t).
\label{Eq:tilde_eps}
\ee
The quasi-classical Green's function $g_{\bf n}$ can then be introduced as
\begin{equation}
g_{\bf n}({\bf r},t,\varepsilon)=\frac{i}{\pi}\int d\xi\;G\left({\bf r},{\bf n}\left( p_{\tilde{\varepsilon}}+\frac{\xi}{v_{\tilde{\varepsilon}}}\right),t,\varepsilon\right).\label{Eq:qcgreens}
\end{equation}
In this equation ${\bf n}={\bf p}/p$ specifies the momentum direction and $v_{\tilde{\varepsilon}}=p_{\tilde{\varepsilon}}/m$. In order to derive an equation for $g_{\bf n} ({\bf r},t)$, one should first consider the difference of Eq.~(\ref{eq:greens}) and its conjugate equation
\be
&&G(\bfr_1,t_1,\bfr_2,t_2)(-i\partial_{t_2}-\hat{\eps}_{2}-\vartheta(\bfr_2,t_2))\\
&&-\int dt_3 \;G(\bfr_1,t_1,\bfr_2,t_3)\Sigma(\bfr_2,t_3,t_2)=\delta_{\bfr_1\bfr_2}\delta_{t_1t_2}.\no
\ee
The result can be written as
\be
&&\left(i\partial_t+\frac{i}{m}{\bf p}\nabla \right)G({\bf r},{\bf p},t,\varepsilon)-[\vartheta({\bf r},t)\stackrel{\bullet}{,}G({\bf r},{\bf p},t,\varepsilon)]\no\\
&=&[\Sigma({\bf r},t,\varepsilon)\stackrel{\bullet}{,}G({\bf r},{\bf p},t,\varepsilon)].
\ee
Here we introduced the $\bullet$-product
\be
&&A({\bf r},{\bf p},t,\varepsilon)\bullet B({\bf r},{\bf p},t,\varepsilon)=\label{Eq:bullet}\\
&&\mbox{e}^{\frac{i}{2}(\nabla^A_{\bf r}\nabla^B_{\bf p}-\nabla^B_{\bf r}\nabla^A_{\bf p}-\partial^A_t\partial^B_\varepsilon+\partial^B_t\partial^A_\varepsilon)}
A({\bf r},{\bf p},t,\varepsilon)B({\bf r},{\bf p},t,\varepsilon).\no
\ee
Due to the slowness of $\Sigma$ and $\vartheta$ a gradient expansion can be performed, where we keep the leading terms only. The quasiclassical approach in its original form does not involve an approximation with respect to the time arguments. Here,  we make an additional smoothness assumption. Namely, we assume that the time variation of the density (and thereby of $\vartheta$) is sufficiently slow to justify the neglect of terms of the order of $\partial_t^2 \vartheta$. In addition, the modulus of the momentum ${\bf p}$ multiplying $\nabla$ is set to $p_{\tilde{\epsilon}}$ and the equation integrated in $\xi$,
thereby obtaining an equation for the quasi-classical Green's function,
\be
&&i\partial_tg_{\bf n}(\bfr,t,\eps)+\frac{i}{m}{\bf n}\nabla \left( p_{\tilde{\varepsilon}}\;g_{\bf n}(\bfr,t,\eps)\right).\label{eq:eil}\\
&&-\frac{i}{p_{\tilde{\varepsilon}}}\nabla \vartheta(\bfr,t)\partial_{\bf n}g_{\bf n}(\bfr,t,\eps)+i\partial_t\vartheta(\bfr,t)\partial_{\varepsilon} g_{\bf n}(\bfr,t,\eps)\no\\
&&+\frac{i}{2\tau_{\tilde{\eps}}}[\left\langle g_{\bf n}(\bfr,t,\eps)\right\rangle_\bfn,g_{\bf n}(\bfr,t,\eps)]\no\\
&&=i[f(\bfr,t,\eps)\sigma_{+},g_{\bf n}(\bfr,t,\eps)]\no
\ee
In this formula, $\left\langle(\dots)\right\rangle$
 denotes angular averaging and $\partial_\bfn=\nabla_\bfn-\bfn$ where $\nabla_\bfn$ is defined through the relation
 \be
 \nabla_\bfp=\bfn\partial_p+\frac{1}{p}\nabla_\bfn.
\ee
The following relation for the disorder-part of the self-energy was employed
\be
 \Sigma_{dis}(\bfr,t,\eps)&=&\gamma\int d\eps_\bfp \nu(\eps_\bfp)\left\langle G(\bfp,\bfr,t,\eps)\right\rangle_\bfn\\
&\approx& -i\pi\nu(\tilde{\eps})\gamma g_0(\bfr,t,\eps)\equiv-\frac{i}{2\tau_{\tilde{\eps}}}g_0(\bfr,t,\eps).\no
\ee
The last relation serves as a definition of the scattering rate $\tau_\eps^{-1}=2\pi\nu(\eps)\gamma$ in our model.
It was used that the Green's functions has a peak for  $\eps_\bfp=\tilde{\eps}$, compare the related discussion in Sec.~\ref{sec:noninteracting}.

Equation (\ref{eq:eil}) does not fully determine the Green's function $g_\bfn$. In the quasiclassical approximation, the condition
\be
g_\bfn(\bfr,t,\eps)\mbox{e}^{-\frac{i}{2}(\overleftarrow{\partial}_t\overrightarrow{\partial}_\varepsilon-\overleftarrow{\partial}_\eps\overrightarrow{\partial}_t)}g_\bfn(\bfr,t,\eps)=1.
\ee
is therefore introduced. Keeping terms that result from the expansion of the exponential in this formula, however, exceeds the accuracy of our approximation. We therefore use the constraint in the form
\be
g^2_\bfn(\bfr,t,\eps)=1.
\ee
It can be seen that this constraint is consistent with the time evolution described by Eq. (\ref{eq:eil}). Indeed, when multiplying equation (\ref{eq:eil}) from the left by $g_\bfn$ and adding the result  to the equation that is obtained by first multiplying Eq.~({\ref{eq:eil}) by $g_\bfn$ from the right, one obtains an equation for $g^2_\bfn$. The resulting equation
\be
&&\partial_t g^2_\bfn(\bfr,t,\eps)+\partial_t\vartheta(\bfr,t)\partial_\eps g^2_{\bfn}(\bfr,t,\eps)\no\\
&&+v_{\tilde{\eps}}\bfn\nabla g_\bfn^2(\bfr,t,\eps)-\frac{1}{p_{\tilde{\eps}}}\nabla\vartheta(\bfr,t)\nabla_\bfn g_{\bfn}(\bfr,t,\eps)\no\\
&&-\frac{1}{2\tau_{\tilde{\eps}}}[\left\langle g_{\bfn}(\bfr,t,\eps)\right\rangle_\bfn,g_{\bfn}^2(\bfr,t,\eps)]\no\\
&&=-[f(\bfr,t,\eps)\sigma_+,g^2_{\bfn}(\bfr,t,\eps)]
\ee
is solved by $g^2_\bfn(\bfr,t,\eps)=c$, where $c$ is an arbitrary constant. This constant can be determined in the noninteracting case, where the relations $g^R(\bfr,t,\eps)=1$ and $g^A(\bfr,t,\eps)=-1$ imply $c=1$. It is usually argued \cite{Kopnin01, Larkin68, Shelankov85} that this constraint carries over to the interacting theory, and we will follow this route here.

Equation (\ref{eq:eil}), the analog of the Eilenberger equation in our problem, can be further simplified in the diffusive regime. This reduction will be discussed next. Let us denote by $q$ and $\omega$ the small momenta and frequencies related to the space and time variation of $n_{\varepsilon}$. In the diffusive regime the inequalities $\tau\overline{v}q\ll 1$ and $\omega\tau \ll 1$ are fulfilled for typical velocities $\overline{v}$. If we additionally demand $\tau\overline{v}q\;\vartheta/\tilde{\varepsilon}\ll 1$ and $\omega\tau \;\vartheta/\tilde{\varepsilon}\ll 1$, the main contribution comes from the zeroth angular harmonic of the quasiclassical Green's function. It is worth noting that the expansion is performed assuming that gradients and time derivatives of the potential are small, i.e. it is not an expansion in the strength of $\vartheta$. We can take into account the influence of higher harmonics approximately
 with the help of the ansatz
\be
 g_{\bf n}=g_0+{\bf n} {\bf g},
\ee
where $g_0=\left\langle g_{\bf n} \right\rangle$ and ${\bf g}=d\; {\bf n}\left\langle {\bf n'} g_{\bf n'}\right\rangle$ in $d$ spatial dimensions and ${\bf n}{\bf g}$ is a small perturbation in the diffusive regime.
In this limit, the constraint $g_\bfn^2=1$ results in the condition $1=g_0^2+\{\bfn\bfg,g_0\}$, so that upon integration in $\bfn$ one obtains the relation $g_0^2=1$ as well as $\bfg=-g_0\bfg g_0$.

In order to derive Eq.~(\ref{eq:fundam}), we first integrate Eq.~(\ref{eq:eil}) with respect to $\bfn$. The result is
\be
&&i\partial_tg_0(\bfr,t,\eps)+\frac{i}{dm}\nabla (p_{\tilde{\eps}}\;\bfg(\bfr,t,\eps))\label{eq:g0}\\
&&-i\nabla\vartheta(\bfr,t)\frac{1}{p_{\tilde{\varepsilon}}}\frac{d-2}{d}\bfg(\bfr,t,\eps)+i\;\partial_t\vartheta(\bfr,t)\;\partial_\eps g_0(\bfr,t,\eps)\no\\
&&=-i[f\sigma_+,g_0(\bfr,t,\eps)].\no
\ee
In a second step we first multiply Eq.~(\ref{eq:eil}) by $\bfn^i$ before integrating in $\bfn$ and find
\be
&&i\partial_t\bfg(\bfr,t,\eps)+\frac{i}{m}\nabla (p_{\tilde{\eps}}\;g_0(\bfr,t,\eps))\\
&&+\frac{i}{p_{\tilde{\eps}}}\nabla\vartheta(\bfr,t)\;g_0(\bfr,t,\eps)
+i\;\partial_t\vartheta(\bfr,t)\;\partial_\eps\bfg(\bfr,t,\eps)\no\\
&&+\frac{i}{2\tau_{\tilde{\eps}}}[g_0(\bfr,t,\eps),\bfg(\bfr,t,\eps)]=-i[f\sigma_{+},\bfg(\bfr,t,\eps)]\no
\ee
After multiplying this equation by $g_0$ from the left and using the relation $g_0[g_0,\bfg]=2\bfg$, we can formally solve for $\bfg$. Due to the smallness of $\bfg$, not all terms need to be kept, and we may work with
\be
\bfg(\bfr,t,\eps)&=& -\frac{\tau_{\tilde{\eps}}}{m} g_0(\bfr,t,\eps) \nabla  (p_{\tilde{\eps}}\; g_0(\bfr,t,\eps))\no\\
&&-\frac{\tau_{\tilde{\eps}}}{p_{\tilde{\eps}}}\nabla\vartheta(\bfr,t,\eps)\no\\
&=&-l_{\tilde{\eps}}g_0(\bfr,t,\eps)\nabla g_0(\bfr,t,\eps),
\ee
where $l_{\tilde{\eps}=}v_{\tilde{\eps}}\tau_{\tilde{\eps}}$. We plug this expression for $\bfg$ into Eq.~(\ref{eq:g0}).  In this way, we obtain the following equation for $g_0$
\be
&&\partial_tg_0(\bfr,t,\eps)+\partial_t\vartheta(\bfr,t)\partial_\varepsilon g_0(\bfr,t,\eps)\\
&&-(\nabla+\Gamma_{\tilde{\eps}}\nabla\vartheta(\bfr,t))(D_{\tilde{\eps}} g_0(\bfr,t,\eps)\nabla g_0(\bfr,t,\eps))\no\\
&&=-[f(\bfr,t,\eps)\sigma_{+},g_0(\bfr,t,\eps)],\no
\ee
where ${D}_{\tilde{\eps}}=\tilde{v}_{\tilde{\eps}}^2\tau_{\tilde{\eps}}/d$. We used the relation $l_{\tilde{\eps}}/p_{\tilde{\eps}}\times(2-d)/d=\Gamma_{\tilde{\eps}}D_{\tilde{\eps}}$, where we defined the quantity
\be
\Gamma_{\tilde{\eps}}=-\partial_\eps\ln \nu(\tilde{\eps})\label{eq:Gamma}.
\ee
$\Gamma$ vanishes in two spatial dimensions, since the density of states is constant. In three dimensions, however, $\Gamma=\frac{2-d}{2\eps}$ is finite. As mentioned before, the above equation should be supplemented with the matrix constraint $g_0^2(\bfr,t,\eps)=1$.

Before making a specific ansatz for the solution, let us focus on the function $f$ that specifies the injection of the wave-packet and initial evolution up to times of the order of the scattering time $\tau$, see Eq.~(\ref{eq:f}). If $F({\bf p},{\bf r})$ is sufficiently smooth in the sense that for typical $v=p/m$ and $q$ controlled by $F({\bf p},{\bf q})=\Psi_0({\bf p}+{\bf q}/2)\Psi_0^*({\bf p}-{\bf q}/2)$ the inequality $\tau v q\ll 1$ holds, we can approximately replace
\be
&&2\pi\nu_{\tilde{\eps}}\;f({\bf r},t,\varepsilon)\\
&\approx&\delta(t)\int (d\bfp)F({\bf p},{\bf r})2\pi\delta\left(\varepsilon_{\bf p}+\vartheta({\bf r},0)-\varepsilon\right)\no\\
&\equiv& \delta(t)\;F(\varepsilon-\vartheta({\bf r},0),{\bf r})\no\label{Eq:Distr}.
\ee
Next we introduce the following ansatz for $g_0$:
\be
g_0(\bfr,t,\eps)=\left(\ba{cc}1&\frac{1}{\pi\nu(\tilde{\eps})}\tilde{n}(\bfr,t,\eps)\\0&-1\ea\right),
\ee
which solves the equation provided $\tilde{n}$ fulfills the kinetic equation
\be
&&\partial_t\tilde{n}(\bfr,t,\eps)-\nabla(D_{\tilde{\eps}}{\nabla}_{\Gamma}\tilde{n}(\bfr,t,\eps))\no\\
&&+\partial_t\vartheta(\bfr,t)\partial_\eps\tilde{n}(\bfr,t,\eps)=\delta(t)2\pi\nu(\tilde{\eps}) F(\tilde{\eps},\bfr).
\ee
Here, we used the notation $\nabla_\Gamma=\nabla-\nabla\vartheta(\bfr,t)\Gamma_{\tilde{\eps}}$. This concludes our derivation of the kinetic equation from Eqs.(\ref{eq:greens}). The diagrammatic interpretation of the different terms appearing in this equation was provided in Sec.~\ref{sec:intperturbation} for the two-dimensional case. The main new ingredient for $d\ne 2$ is the non-constant density of states. Within our model, it results in a frequency-dependent scattering time [compare Eqs. (\ref{eq:Gamma})]. Since the density of states enters with argument $\tilde{\eps}=\eps-\vartheta(\bfr,t)$, the disorder part of the self-energy $\Sigma_{dis}$ explicitly depends on $\vartheta$. In a diagrammatic language, it means that a generalization of the box diagrams $\mathcal{B}$ is required for a non-constant density of states in order to accommodate this change, see Fig.~\ref{fig:BRBA3d}. This modification was first noticed in Ref.~\onlinecite{Cherroret11}.

\begin{figure}[h]
\setlength{\unitlength}{2.3em}
\includegraphics[width=8\unitlength]{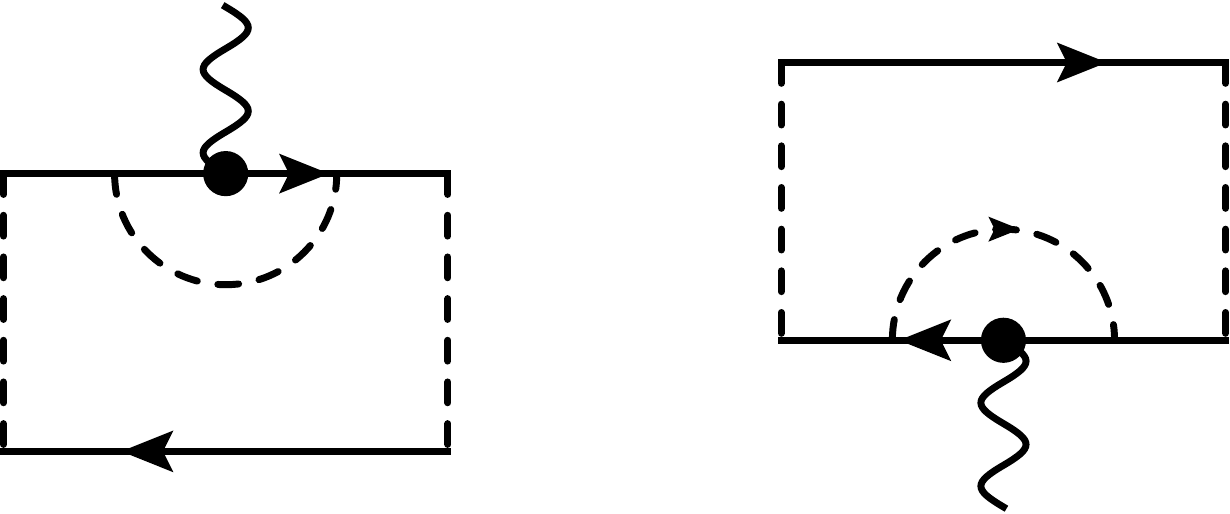}
\caption{For dimension $d\ne 2$ the density of states is not constant and the $\vartheta$-dependence of the Green's function entering the SCBA becomes important. In this case the box diagrams $\mathcal{B}_R$ and $\mathcal{B}_A$ should be generalized as displayed. \cite{}}
\label{fig:BRBA3d}
\end{figure}

One may write the distribution function as a function of the kinetic energy instead of the total one, $n({\bf r},\varepsilon,t)=\tilde{n}({\bf r},\varepsilon+\vartheta({\bf r},t),t)$, see Eq.~(\ref{Eq:basic22}). From a technical point of view, this transformation amounts to a gauge transformation. We could have utilized this transformation already at the beginning of our derivation by working with the so-called gauge-invariant Green's function $\mathcal{G}$, which can be introduced as
\be
G(\bfr_1,\bfr_2,t_1,t_2)&=&\int (d\bfp)(d\eps)\;\mathcal{G}(\bfr,\bfp,t,\eps)\no\\
&&\mbox{e}^{i\bfp(\bfr_1-\bfr_2)-i[\eps-\vartheta(\bfr,t)](t_1-t_2)}.\;
\ee
A derivation based on $\mathcal{G}$ instead of $G$, but otherwise following the same lines as described in this section, leads directly to Eq.~(\ref{Eq:basic22}) instead of Eq.~(\ref{eq:fundam}).

\section{Collisions induced by the nonlinearity}
\label{Sec:collisions}

The purpose of the paper is to present the technical aspects of the derivation of the kinetic equation describing
the pulse propagation in a disordered and nonlinear medium. The obtained equation (\ref{Eq:basic22}) describes the diffuse propagation (as a result of collisions with elastic defects) in the self-consistent potential, but so far fully ignores collisions induced by the nonlinearity. With respect to the nonlinearity, this equation describes the collisionless regime.

We now wish to discuss the role of collisions. To this end let us first recall the general spirit of the derivation presented for the collisionsless regime. As is typical for disordered systems, the physics at long time scales and long distances is dominated by diffusion modes. The nonlinearity leads to an interaction of these modes.
To treat this effect, we singled out pairs of fields $\phi$ and $\overline{\phi}$ with a small momentum difference in the interaction term. Afterwards, the effect of interaction of the diffusion modes was considered in a self-consistent way by introducing the smooth classical potential $\vartheta$ as described in Sec.~\ref{Sec:slow}. This procedure may be viewed as a mean field approximation. It should be noted, however, that in this procedure only a small (albeit important) subset of all possible scattering processes was singled out and treated non-perturbatively as a result of self-consistency.
It is important that the potential $\vartheta$ is proportional to the density and therefore smooth and slowly varying. This is the reason why the self-consistent part of the problem of the propagation of the diffusion modes may be treated within the quasiclassical formalism.

To incorporate collisions, one has to go beyond the scheme discussed above. Collisions induced by a nonlinear interaction in classical wave systems are routinely studied in nonlinear physics (see e.g., V. E. Zakharov, V. S. L'vov, and G. Falkovich "Kolmogorov Spectra of Turbulence"). There it works as follows. In the equation of motion for the occupation numbers $n_\bfp(\bfr,t)$, one obtains nonlinear terms, which are considered using the random-phase approximation. At second order in the coupling constant $\lambda$ this procedure yields a collision integral, which in the case of the four-wave interaction (like in the NLSE or GPE) is proportional to the third power in the occupation numbers.

Here we will show how the derivation of the collision integral in the kinetic equation can be obtained in the framework of the field-theoretical approach we use. In order to account for collisions, we have to go beyond the mean-field description employed in the collisionless regime, namely, we need to include fluctuations. We will derive these at the second order with respect to $\lambda$, the lowest order at which collisions appear in the theory. We therefore need to calculate second-order corrections to the self energy. When doing so, we will assume that the diffusive propagation in the field of the self-consistent potential created by the nonlinearity is already known according to the analysis presented in the previous section.

In the calculation of the collision integral, we will use the Green functions $G(\bfr_1,t_1;\bfr_2,t_2)$ as defined in Eq.~(\ref{eq:G1}). When doing so, we neglect terms containing $F$, cf. Eq.~(\ref{eq:shift1}), because $F$ decays on a scale of the mean-free path in the disordered medium. The off-diagonal component $G_{12}$, in turn, describes the long-range nonlinear diffusion of a partial wave until the moment of collision with another partial wave at $t\approx t_1\approx t_2$. The component $G_{12}$ resembles the Keldysh component in the regular technique. It is non-vanishing due to the injection process at $t=0$, which is encoded in the source term $S_s$ of the action. We will therefore denote $G_{12}$ as $G^{S}$.

Since $S_{int}$ originates from the NLSE/GPE, the theory used in this paper contains only classical vertices, namely those that couple one of the quantum components of the doublets $\phi$ or $\overline{\phi }$ with three classical ones, compare Fig.~\ref{fig:blocks1}. As a consequence of this fact, there are only three contributions (diagrams) to be calculated for self-energies: one for $\Sigma^{S}$ and the other two for each of the diagonal components, e.g., for $\Sigma^{R}$. As a result one obtains

\begin{equation}
\Sigma^{S}(x_1,x_2)=-2\lambda ^{2}G^{S}(x_2,x_1)G^{S}(x_1,x_2)G^{S}(x_1,x_2),
\label{Eq:SigmaS}
\end{equation}

and

\be
&&\Sigma^{R}(x_1,x_2)=-2\lambda ^{2}[G^{A}(x_2,x_1)G^{S}(x_1,x_2)G^{S}(x_1,x_2)\no\\
&&+2G^{S}(x_2,x_1)G^{R}(x_1,x_2)G^{S}(x_1,x_2)].
\label{Eq:SigmaR}
\ee

\begin{figure}
\setlength{\unitlength}{2.3em}
\includegraphics[width=11.5\unitlength]{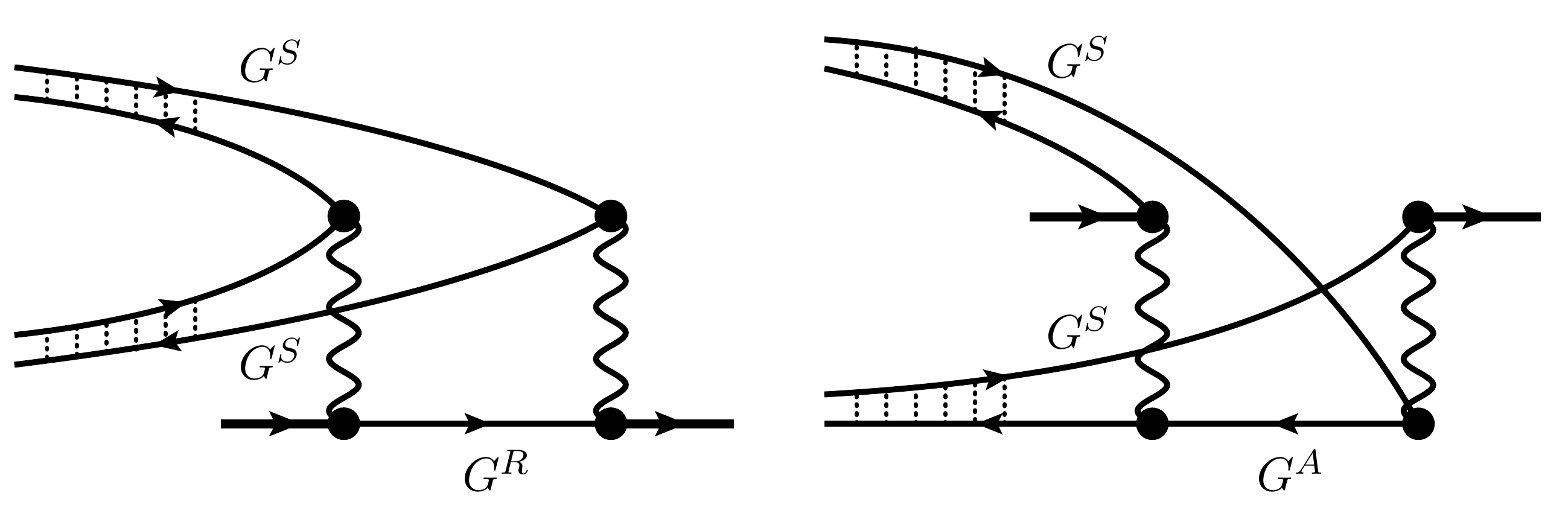}
\caption{Corrections to $\Sigma^R$ according to Eq.~\ref{Eq:SigmaR}.}
\label{fig:Sigma_R}
\end{figure}

\begin{figure}
\setlength{\unitlength}{2.3em}
\includegraphics[width=6\unitlength]{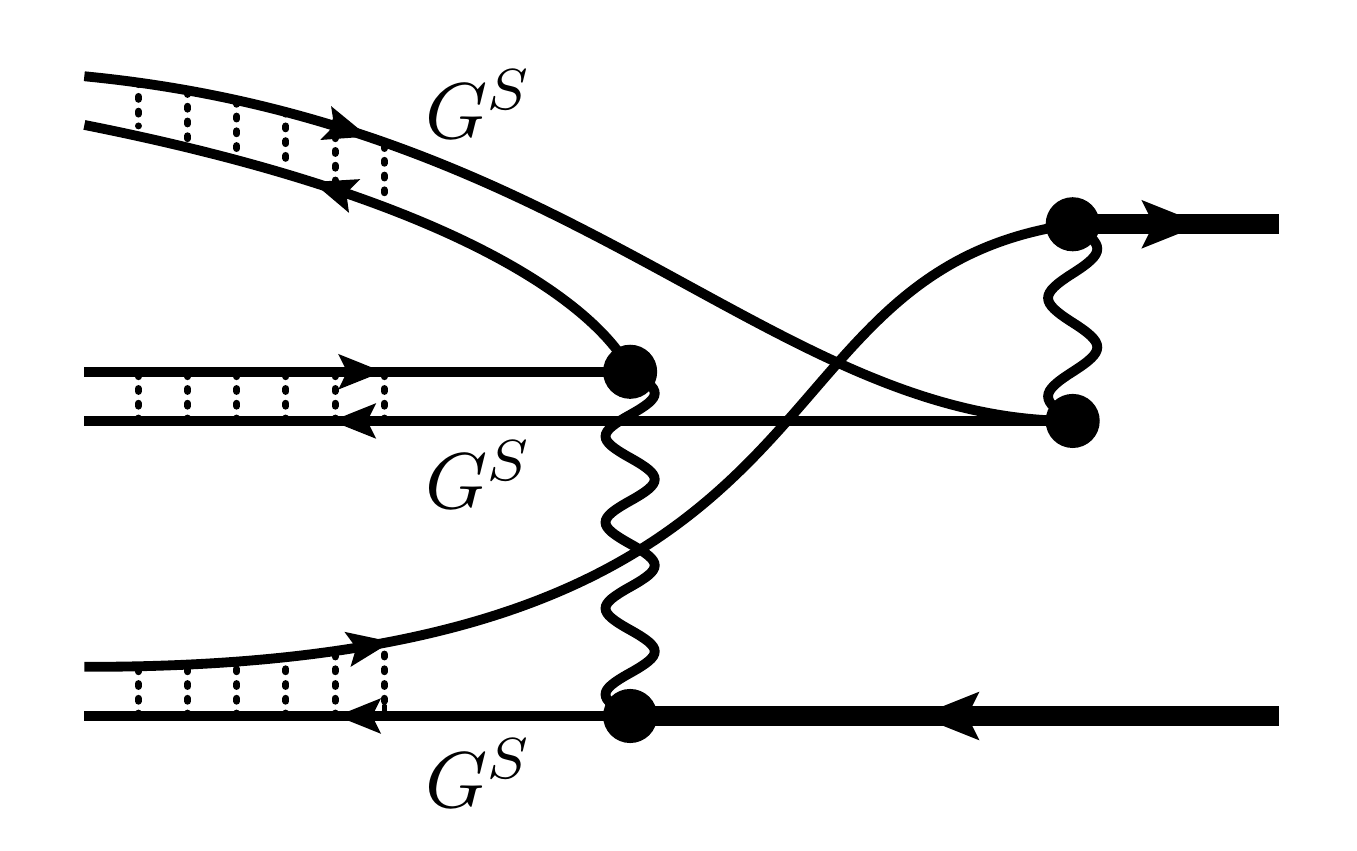}
\caption{Corrections to $\Sigma^S$ according to Eq.~\ref{Eq:SigmaS}.}
\label{fig:Sigma_S}
\end{figure}

These two quantities determine the collision integral in the kinetic equation. In standard kinetic theory, the kinetic equation is formulated in terms of the mass-shell distribution function $n_\bfp(\bfr,t)$. It can be introduced as follows. First, one parametrizes $G^S=G^R\bullet \hat{n}-\hat{n}\bullet G^A$, where $\hat{n}=\hat{n}(\bfr,\bfp,t,\eps)$ and the $\bullet$-product has been defined in Eq.~(\ref{Eq:bullet}). Then one defines the on-shell distribution function as
\be
n_\bfp(\bfr,t)=\hat{n}\left(\bfr,\bfp,t,\eps=\eps_p+\vartheta+\Re(\Sigma^R)\right).
\label{Eq:mass-shell}
\ee
This definition is motivated by the observation that as long as $G^S$ is a smooth function of coordinates and times, the largest contribution to $G^S$ comes from the product of Wigner transforms
\be
G^S(\bfr,\bfp,t,\eps)&\approx& \hat{n}(\bfr,\bfp,t,\eps)(G^R-G^A)(\bfr,\bfp,t,\eps)\label{Eq:GS}\\
&\approx& -2\pi i \hat{n}(\bfr,\bfp,t,\eps)\tilde{\delta}(\eps-\eps_p-\vartheta-\Re(\Sigma^R))\no\\
&\approx& -2\pi i n_\bfp(\bfr,t)\delta(\eps-\eps_p-\vartheta-\Re(\Sigma^R)).\no
\ee
Here, $\tilde{\delta}$ is a broadened $\delta$-function, which is sharply peaked compared to the scale of variation of $\hat{n}$ and can therefore be replaced by a regular delta function. Since the distribution function $\hat{n}$ always appears in combination with the delta function, it is useful to work with the mass-shell distribution function $n_\bfp$ defined in Eq.~(\ref{Eq:mass-shell}).

In a general context, $\vartheta$ could be an external potential.
For our application, an external potential is not present, but we treat a part of $\Re{\Sigma}^R$ separately, namely the self-consistent potential $\vartheta(\bfr,t)=2\lambda n(\bfr,t)$. Therefore, in Eq.~(\ref{Eq:GS}) $\Re{\Sigma}^R$ should be understood as the real part of the $\Sigma^R$ as given in Eq.~(\ref{fig:Sigma_R}). As will be discussed further below, in the regime of applicability of our approach $\Re{\Sigma}^R$ may be considered to be small compared to $\vartheta$, and we will not mention it further.

In an approximation consistent with this reasoning, the collision integral can be written as
\be
&&\hat{I}^{coll}(\bfr,{\bf p},t,\eps)\label{Eq:Coll Int}
\\
&=&i\Sigma^{S}(\bfr,\bfp,t,\eps)+2\hat{n}(\bfr,\bfp,t,\eps) \Im\Sigma^{R}(\bfr,\bfp,t,\eps).\no
\ee
Entering with Eq.~(\ref{Eq:GS}) into the expressions (\ref{Eq:SigmaS}) and (\ref{Eq:SigmaR}), and replacing the Wigner transform of the products of Green's functions on the RHS of both equations by the product of Wigner transforms, one finds from Eq.~(\ref{Eq:Coll Int}) the collision integral for the mass-shell distribution function
\be
&&I^{\rm coll}(\bfr,\bfp,t)=\hat{I}(\bfr,\bfp,t,\eps=\eps_p+\vartheta)\\
&=&\frac{4\pi\lambda^2}{(2\pi)^{2d}}\int d\bfp_2d\bfp_3d\bfp_4\no  \\
&&\delta (\bfp+\bfp_2-\bfp_3-\bfp_4) \delta(\eps_p+\eps_{p_2}-\eps_{p_3}-\eps_{p_4})\no\\
&&\times \{[n_\bfp+n_{\bfp_2}]n_{\bfp_3} n_{\bfp_4}-n_{\bfp}n_{\bfp_2}[n_{\bfp_3}+n_{\bfp_4}]\},\no
\label{Eq:Coll}
\ee
where we suppressed  (for the sake of brevity) the space and time arguments $x=({\bf r},t)$ in the distribution functions. The normalization is $n(\bfr,t)=\int (d\bfp) n_\bfp(\bfr,t)$.
One may check that the obtained collision term coincides with the $C_{22}$-term in the kinetic theory of the Bose gas,\cite{Griffin09} the Bose-factors $n_{\bfp}+1$ are replaced by $n_{\bfp}$. Most important is that the obtained collision integral is universal, i.e., it does not depend on the system microscopy and holds for both NLS and GP equations.

In this paper, we work with the $\xi$-integrated quasiclassical Green's function and, correspondingly, with a frequency-dependent distribution function rather than with a distribution function depending on the quasiparticle energy $\eps(\bfp)$.  To make a connection, we note that in the equation for $G^S$, Eq.~({\ref{Eq:GS}}), where the $\delta$-function was used to fix the frequency argument $\eps$ of $n$, we may alternatively fix $\eps_p$ and thereby the modulus of $\bfp$
\be
G^S(\bfr,\bfp,t,\eps)\approx -2\pi i \hat{n}(\bfr,p_{\tilde{\eps}}\bfn,t,\eps)\delta(\tilde{\eps}-\eps_p),
\ee
where $\tilde{\eps}$ is defined in Eq.~(\ref{Eq:tilde_eps}). With the help of this representation one finds
\be
&&I^{\rm coll}_{\bfn}(\bfr,t,\eps)\equiv\hat{I}^{coll}(\bfr,p_\eps \bfn,t,\eps+\vartheta(\bfr,t))\\
&&=4\pi \lambda^2 (2\pi)^d \int d\bfn_2d\bfn_3d\bfn_4\int  d\eps_2 d\eps_3 d\eps_4 \no\\
&&\times \nu(\eps_2) \nu(\eps_3)\nu(\eps_4) \delta(\eps+\eps_2-\eps_3-\eps_4)\no\\
&&\times \delta(\bfp_\eps +\bfp_{\eps_2}-\bfp_{\eps_3}-\bfp_{\eps_4})\Big([n'_{\bfn,\eps}+n'_{\bfn_2,\eps_2}]n'_{\bfn_3,\eps_3}n_{\bfn_4,\eps_4}'\no\\
&&-n_{\bfn,\eps}'n_{\bfn_2,\eps_2}'[n_{\bfn_3,\eps_3}'+n_{\bfn_4,\eps_4}']\Big),\no
\ee
where $\bfp_{\eps_i}=p_{\eps_i}\bfn$ and we suppressed the space and time arguments in the distribution functions $n_{\bfn,\eps}'(\bfr,t)=\hat{n}(\bfr,p_\eps\bfn,t,\eps+\vartheta(\bfr,t))$.  Only positive values of $\eps_i$ are included in the integration.

In the diffusive limit, which we concentrate on in this paper, only  the isotropic part of $n'_{\bfn}$ is important and $n'_{\bfn}$ may be replaced by its angular average $n'=\int d\bfn\;n'_{\bfn}$. In a similar way, the knowledge of $I^{\rm coll}=\int d\bfn\; I^{\rm coll}_{\bfn}$ is sufficient. In accordance with Ref.~\onlinecite{Schwiete10}, the normalization of the distribution function $n(\bfr,t,\eps)$ in Sec.~\ref{sec:quasiclassics} has been chosen such that $n(\eps)=2\pi\nu(\eps)n'(\eps)$. The resulting full kinetic equation including the collision integral is written in Sec.~\ref{Sec:discussion},  Eq.~(\ref{Eq:Fullequation}).

In the course of derivation of the kinetic equation we omitted the renormalization induced by the real part of $\Sigma^R$. This kind of renormalization is standard for any many-body problem. The corrections induced by the real part, for example $\partial_{\eps}\Re{\Sigma^R}$, are of the order of $(\lambda n/\overline{\eps})^2$, and are therefore smaller than the leading terms which are kept in the kinetic equation.

\section{Conclusion}
\label{Sec:conclusion}

In this work, we discussed the propagation of a wave-packet in a disordered and nonlinear medium, for which the dynamics is governed by the NLSE/GPE. Possible applications include the propagation of a light beam in a nonlinear optical medium and the expansion of a cloud of Bose atoms released from a trap. For definiteness, we use the term "particles" irrespective of the system.

We considered the case when the potential (interaction) energy induced by the nonlinearity is considerably smaller than the typical kinetic energy. This allowed us to use the picture of a gas of particles moving in a self-consistent potential rather than that of a hydrodynamic flow. Diffusion occurs as a result of elastic scattering from a random potential. We studied a regime for which particles scatter on impurities many times before colliding with other particles.

Another important consequence of the smallness of the nonlinearity is the possibility to neglect off-diagonal terms in the Bogoliubov transformation. In the case of a Bose condensate released from a trap, our consideration corresponds to a stage of evolution when the initial hydrodynamic flow\cite{Kagan96,Castin96} of the Bose atoms already passed by and particles diffuse with a typical kinetic energy of the order of the (initial) chemical potential and the wavelength $\lambda_{typ}$ comparable with the healing length $\xi$ of the trapped condensate. We assume that $\lambda_{typ}$ is much shorter than the mean free path. Since we use the GPE, which arises as the classical equation of motion in the theory of the interacting Bose gas, it is assumed that the occupation numbers $n_\bfp$ with $p\sim 2\pi/\lambda_{typ}$ remain large on the discussed stage of the expansion.

Compared to  the case of disordered electrons with electron-electron interactions\cite{Altshuler85} virtual processes involving diffusion modes need not be considered for the discussed problem. As we have already mentioned, such processes give corrections that are small in the parameter $1/\overline{\eps}\tau\ll 1$, but unlike for electrons at low temperature, they are not accompanied with non-analytic corrections, which make them important in the case of the degenerate electron gas.

In the case of two-dimensional particles, $d=2$, the motion in the plane is not constraint, while the third dimension either represents the effective time-like direction in the case of optics experiments or is blocked by the quantization induced by a potential that restricts the motion in the transverse direction. It is important to distinguish the original dimension of the single particle states in the NLSE/GPE, denoted with $d$ in this paper, from the effective dimension of the diffusive collective modes, which may be different. Namely, the derived kinetic equation can be solved in different geometries.

To illustrate the role of the effective dimensionality, let us consider the example of a stripe made out of $2d$-particles. Then, the diffusion will be described by a one-dimensional solution while particles can be two-dimensional if the quantization with respect to the width of the stripe can be neglected. As in the case of $2d$-particles diffusing in a plane, the exact solution for the time-dependence of the mean squared radius still holds. We expect that the existence of this simple analytical result can be useful for numerics or suitably designed experiments.

Besides technical details of the derivation of the two self-consistent equations describing the collisionless regime, the present paper contains a discussion of inter-particle collisions. The collision integral has been obtained from the same field-theoretical approach that was used as a starting point for the derivation of the kinetic equation in the collisionless regime and the procedure was straightforward. It is important to stress that the collision integral is the same for both optics (NLSE) and cold atoms (GPE), i.e., independent of the microscopic origin. Since the inter-particle collisions are elastic and local, it does not alter the relation (\ref{eq:dtrsq}) between $\left\langle r^2\right\rangle$ and $t$ in the case of a constant density of states. The result remains valid as long as the kinetic equation is applicable, despite the fact that the collisions change the dynamics of propagation.

The change of dynamics caused by collisions is qualitatively different from the one introduced by the self-consistent potential. Indeed, the smooth self-consistent potential is responsible for a gradual change of the kinetic energy during the expansion. In contrast, the energies of incoming and outgoing particles participating in a collision process may differ considerably. In particular, in three dimensions the collision-induced redistribution of energies may lead to a population of localized particles with energies $\eps\lesssim 1/\tau$. This mechanism bears a certain similarity with the seeding of a macroscopic occupation of low-energy states in a trapped Bose gas; this step is crucial for the formation of a Bose-condensate starting from a confined Bose gas \cite{Kagan92, Kagan95, Semikoz97}. In the case of the expanding disordered Bose gas in $3d$, collisions seed a population of particles that are likely to localize. An estimate for the rate of generation of localized particles may be obtained from the in-scattering term of the collision integral upon integration over the interval $0<\eps<1/\tau$, namely
 \be
dn_{loc}/dt\approx dn/dt|_{\varepsilon \lesssim 1/\tau }\sim \frac{1}{\tau _{coll}}\frac{n'(\varepsilon \sim 2\overline{\varepsilon })}{n'(\varepsilon \sim \overline{\varepsilon })}\frac{n(t)}{(\tau \overline{\varepsilon })^{d/2}}.
\ee
Here, we assumed that both colliding particles have the energy $\sim \overline{\varepsilon}$. Although the discussed effect is important under static conditions, one may show that for the situation we study the seeding of localized states for an expanding cloud is negligible because of the fast drop of $n(t)$.

The situation in two dimensions is different in that in the absence of interactions all states are localized on the scale of the localization length, $l_{loc}(\eps)$. For a state with the energy $\eps$, the process of localization starts to develop at a time of the order of $l^2_{loc}(\eps)/D(\eps)$. In the presence of interactions, however, this picture changes. First of all, both the time-dependent potential and the interparticle collisions lead to dephasing, which weakens localization effects. One may show that in $2d$ if the number of particles is large enough, the expanding cloud will pass $l_{loc}$ without being stopped. We will discuss this situation in more detail elsewhere.

\begin{acknowledgments}
We thank T.~Wellens, K.~Tikhonov and especially G.~Falkovich for useful discussions.
The authors gratefully acknowledge the support by the Alexander von Humboldt Foundation, and thank the members of the Institut f\"ur Theorie der Kondensierten Materie at KIT for their kind hospitality. G.~S.~ also acknowledges financial support by the Albert Einstein Minerva Center for Theoretical Physics at the Weizmann Institute of Science.  A.~F. is supported by the National Science Foundation grant NSF-DMR-1006752.  
\end{acknowledgments}

\appendix

\section{Distribution function}
\label{app:distribution}

The aim of this appendix is to show how the approximation (\ref{eq:distrappr}) for the distribution function $f$ is obtained. For the sake of completeness, we consider the generalization to the case with interaction. Starting point is the definition of $f$ in Eq.~(\ref{eq:distr}). We introduce times $t=(t_1+t_2)/2$ and $\Delta t=t_1-t_2$ as well as coordinates $\bfr=(\bfr_1+\bfr_2)/2$ and $\bfrho=\bfr_1-\bfr_2$ and write
\be
G(\bfr,\bfp,t,\eps)&=&\int d(\Delta t)d\bfrho\;G(\bfr_1,\bfr_2,t_1,t_2)\;\mbox{e}^{-i\bfp\bfrho+i\eps\Delta t}.\no\\
\ee
Note that the Green's function $G$ depends on coordinates $\bfr$, $t$ only via the self-consistent potential $\vartheta$, since otherwise the disorder averaged system would be translationally invariant in time and coordinates. If variations of $\vartheta$ in time in space are slow in comparison to other relevant scales in the system, we may write the leading term in a gradient expansion for $f(\bfr,t,\eps)$ (defined in Eq.~(\ref{eq:feps})) as
\be
f(\bfr,t,\eps)&\approx&\gamma\int (d\bfp)(d\bfq)\; (d\omega)\;\underline{G}^R\left(\bfr,\bfp_+,t,\eps_+\right)\no\\
&&\times F(\bfp,\bfq) \mbox{e}^{i\bfq\bfr} \underline{G}^A\left(\bfr,\bfp_-,t,\eps_-\right)\mbox{e}^{-i\omega t}.\;
\ee
Both Green's functions decay on typical time scales of the order of the mean free path. If $f$ is convoluted with a function that is smooth of this time scale, we may therefore use $\overline{f}(\bfr,t,\eps)\approx \delta(t)\int_{-\infty}^{\infty}dt f(\bfr,\eps,t)$ instead. Next, it is assumed that $F$ controls momenta $\bfq$ so that essential $q=|\bfq|$ are small as $ql_{typ}\ll 1$. Then,
\be
&&\overline{f}(\bfr,t,\eps)\approx \delta(t)\int (d\bfp)(d\bfq)\frac{\gamma F(\bfp,\bfq)\mbox{e}^{i\bfq\bfr}}{[\eps-\eps_{\bfp}-\vartheta(\bfr,0)]^2+\frac{1}{(2\tau_{\tilde{\eps}})^2}}\no\\
\ee
The Lorentzian is peaked around $\eps\sim \eps_\bfp+\vartheta(\bfr,0)$ and has a width of the order of $\tau$.

If the distribution function is used to average a quantity that depends smoothly on $\eps$ as is the case for our problem (where $\eps$ determines the diffusion coefficient), the Lorentzian acts essentially like a smeared $\delta$ function and we may use $
\overline{\overline{f}}(\bfr,t,\eps)=\tau_{\tilde{\eps}}(2\pi)\delta(\eps-\eps_\bfp-\vartheta(\bfr,0))\int (d\eps) \overline{f}(\bfr,t,\eps).$
This leads us to the result of this appendix,
\be
&&2\pi\nu_{\tilde{\eps}} \overline{\overline{f}}(\bfr,t,\eps)=\\
&&\delta (t)\;\int (d\bfp)\;F(\bfp,\bfr)\;(2\pi)\delta\left(\eps_\bfp+\vartheta(\bfr,0)-\eps\right)\no
\ee
We used the relation $\gamma\tau_{\tilde{\eps}}=1/(2\pi\nu_{\tilde{\eps}})$.  If the smoothness assumptions outlined in this appendix are met, the representation of the distribution function $f$ in the form given by $\overline{\overline{f}}$ is justified. In the noninteracting limit, this leads us to relation (\ref{eq:distrappr}).

\end{document}